%% file: PipelineUniversalGatesPaper.tex
\documentclass[%
aps,
prx,
reprint,
superscriptaddress,
nofootinbib,
amsmath,
amssymb
]{revtex4-2}
\usepackage{
    physics,
    graphicx,
    multirow,
    tabularx,
    mathtools, 
    placeins, 
    nicefrac, 
    hyperref, 
    threeparttable, 
    siunitx,
    enumitem
}

\usepackage[capitalize]{cleveref}
\Crefname{section}{Sec.}{Secs.}

\usepackage[dvipsnames]{xcolor}
\hypersetup{
    colorlinks,
    linkcolor={blue!50!black},
    citecolor={blue!50!black},
    urlcolor={blue!80!black}
}

\usepackage[caption=false]{subfig} 
\graphicspath{{Figures/}}

\begin{document}


\title{Fault-tolerant Quantum Computation without Distillation on a 2D Device}

\newcommand{\qmaddress}{\affiliation{Quantum Motion, 9 Sterling Way, London N7 9HJ, United Kingdom}}
\newcommand{\oxddress}{\affiliation{Department of Materials, University of Oxford, Parks Road, Oxford OX1 3PH, United Kingdom}}

\author{Thomas R. Scruby}
\email{t.r.scruby@gmail.com}
\affiliation{Okinawa Institute of Science and Technology, Okinawa, 904-0495, Japan}

\author{Kae Nemoto}
\email{kae.nemoto@oist.jp}
\affiliation{Okinawa Institute of Science and Technology, Okinawa, 904-0495, Japan}

\author{Zhenyu Cai}
\email{cai.zhenyu.physics@gmail.com}
\qmaddress
\oxddress

\date{\today}

\begin{abstract}
    We show how looped pipeline architectures -- which use short-range shuttling of physical qubits to achieve a finite amount of non-local connectivity -- can be used to efficiently implement the fault-tolerant non-Clifford gate between 2D surface codes described in (Sci. Adv. 6, eaay4929 (2020)). The shuttling schedule needed to implement this gate is only marginally more complex than is required for implementing the standard 2D surface code in this architecture. We compare the resource cost of this operation with the cost of magic state distillation and find that, at present, this comparison is heavily in favour of distillation. The high cost of the non-Clifford gate is almost entirely due to the relatively low performance of the just-in-time decoder used as part of this process, which necessitates very large code distances in order to achieve suitably low logical error rates. We argue that, as very little attention has previously been given to the study and optimisation of these decoders, there are potentially significant improvements to be made in this area. 
\end{abstract}

\maketitle

\section{Introduction}

It is well known that all logical gates implementable by a constant-depth circuit in a two-dimensional topological code must belong to the Clifford group~\cite{bravyiClassificationTopologicallyProtected2013}, and also that any quantum circuit composed entirely of Clifford operations is efficiently simulable on a classical computer~\cite{gottesmanHeisenbergRepresentationQuantum1998}. Other techniques for performing logic in these codes (e.g. defect braiding) are similarly constrained~\cite{Webster_2020}, meaning that more complex and costly procedures must be employed in order to implement the non-Clifford gates necessary for a universal gate set. Typically these procedures involve the preparation of \textit{magic states} which can be used to implement non-Clifford gates via a gate-teleportation circuit. The most commonly considered method of preparing magic states is distillation~\cite{bravyiUniversalQuantumComputation2005}, which uses Clifford operations to transform many low-fidelity copies of a magic state into a small number of higher-fidelity copies. An alternative is to use a three-dimensional topological code where transversal non-Clifford gates can be implemented fault-tolerantly and magic states can be prepared in constant time, but simulations of the performance of this approach suggest that it does not substantially improve space-time overhead~\cite{beverlandCostUniversalityComparative2021} and it also comes with the added requirement of either a three-dimensional architecture or extensive non-local connectivity.

These issues can be avoided if we allow for logical operations implemented by circuits with non-constant depth. Making these operations fault-tolerant is more challenging as it can be difficult to correct for errors which occur at intermediate points in the circuit, but one example that can be made fault-tolerant is a linear-depth (in the code distance) CCZ gate between three 2D surface codes~\cite{brownFaulttolerantNoncliffordGate2020}. This gate implementation works by exchanging a spatial dimension for a temporal one so that the constant time CCZ gate in the 3D surface code~\cite{vasmerThreedimensionalSurfaceCodes2019} becomes a linear-time gate in the 2D code. The fault-tolerance of this procedure comes from a decoding strategy called \textit{just-in-time} (JIT) decoding~\cite{bombin2DQuantumComputation2018} which attempts to guess the syndrome of the corresponding 3D code at each time step of the circuit. The logical gate fidelity will therefore be lower than for the constant-time gate in the 3D code (where no guessing is required) but $O(d)$ fewer physical qubits are necessary for its implementation.

In addition to this non-standard decoding strategy, the linear time gate also requires a bounded amount of non-local connectivity, and so cannot be implemented on a device with strictly 2D connectivity. However, recent hardware developments have shown that this is a much less demanding requirement than it might once have appeared~\cite{ryan-andersonHighfidelityTeleportationLogical2024,bluvsteinLogicalQuantumProcessor2024}, and in particular the looped pipeline architecture in~\cite{caiLoopedPipelinesEnabling2023} uses short-range shuttling of physical qubits to provide exactly this kind of connectivity. 

In this work we show exactly how to use a looped pipeline architecture to implement this gate and examine the corresponding resource cost relative to the cost of using the same architecture for magic state distillation. The physical operations required for this (short-range shuttling on fixed paths, single- and two-qubit physical gates, single qubit measurement in a Pauli basis) are identical to those required for the implementation of multiple copies of the standard planar surface code in this architecture, and the only differences in our case are the inclusion of local, constant-depth circuits that compile the physical CCZ gate and minor modifications to the shuttling sequence. Given that a fault-tolerance proof for the JIT decoder was presented in \cite{brownFaulttolerantNoncliffordGate2020} we thus expect the entire protocol to be fault-tolerant.
We begin by reviewing the linear time CCZ gate and looped pipleline architecture in \cref{section:background}, then give a full description of the necessary error correcting codes and their mappings to this architecture in \cref{section:architecture}. Finally, in \cref{section:resource} we compare the resource cost for both the conventional distillation-based approach and the linear-time CCZ in this architecture.

\section{Background}

In this section we give a high-level overview of the linear-time CCZ gate (\cref{subsection:JIT_background}) and the looped pipeline architecture (\cref{looped_pipeline}). Readers wishing to read about these topics in more detail can consult~\cite{brownFaulttolerantNoncliffordGate2020,scrubyNumericalImplementationJustTime2022} and~\cite{caiLoopedPipelinesEnabling2023}. 

\label{section:background}
\subsection{Surface code without distillation}
\label{subsection:JIT_background}
\input{JIT_background}
\subsection{Looped pipeline}
\label{looped_pipeline}
\input{Looped_pipeline}

\section{Main Architecture}
\label{section:architecture}
\input{architecture}

\section{Resource Comparison}
\label{section:resource}
\subsection{Problem setting}
In this section, we will compare the resource overhead of the architecture with and without magic state factories. We will consider the case of implementing parallel CCZ gates covering all qubits, which is a bottleneck step in the conventional approach with magic state factories if we want to implement the carry lookahead adder in \cite{draperLogarithmicdepthQuantumCarrylookahead2006} (to be exact, it does not require parallel CCZ covering exactly all qubits, but it is a good approximation). Adders are key components in Shor's algorithm, and to factor 2048 bit numbers, we need around $N = 6000$ logical qubits, $D_{\mathrm{cycle}} = 2.5 \times 10^{10}$ code cycles and $M_{\textsc{ccz}} = 3 \times 10^9$ CCZ gates~\cite{gidneyHowFactor20482021}. Hence, our targeted logical error rate per code cycle and our target logical CCZ error rate are $P_{\mathrm{cycle}}  = 1/N/D_{\mathrm{cycle}} \sim 6 \times 10^{-15}$ and $P_{\textsc{ccz}} = 1/M_{\textsc{ccz}} \sim 3 \times 10^{-10}$, respectively. Note that here we are assuming magic state distillation is not the bottleneck when we are calculating $P_{\mathrm{cycle}}$. 

We will be considering the case where the physical gate error is $p = 5\times 10^{-4}$. Using the formula $P_{\mathrm{cycle}} = 0.1 (100 p)^{(d+1)/2}$~\cite{fowlerLowOverheadQuantum2019}, the minimum code distance $d$ we need to achieve the target logical error rate $P_{\mathrm{cycle}} \leq 6 \times 10^{-15}$ is $d = 21$. 

Using the CCZ factories outlined in Ref.~\cite{gidneyEfficientMagicState2019}, fixing the level-2 distance at $d = 21$, the level-1 distance $d_1$ needed to achieve the target logical error rate $P_{\textsc{ccz}} \leq 3 \times 10^{-10}$ is $d_1 = 13$ (outputs $P_{\textsc{ccz}} \sim 10^{-12}$). The footprint of the CCZ factories is shown in Figure 1 of Ref.~\cite{gidneyEfficientMagicState2019}, which is $12d_1 \times (16d_1 + 4d) \approx 7.5d \times 14 d$, for our case. 
The magic state production rate is limited by the level-1 distance in this case since $d_1 * 2 + 1 \geq d$. Hence, the rounds of repetition needed for operations like lattice surgery is $d_1 \times 2 + 1 = 27$, 
and we need $8.5$ rounds, which means the number of code cycles needed to produce one CCZ state is $27 \times 8.5 \approx 230$. If we allow consecutive rounds of magic state production to overlap, we can potentially reduce the number of rounds needed from $8.5$ to $5.5$~\cite{gidneyEfficientMagicState2019}, but we will not make use of this here for simplicity. We need to implement one round of logical CNOT for $T$ gate teleportation, which will take $2d$ code cycles using lattice surgeries. After the teleportation, we require up to $3$ (and on average $1.5$) CZ corrections based on the teleportation measurement result, which adds in $1.5 \times 2d = 3d$ more code cycles. Hence, the total number of code cycles needed to implement a CCZ gate from the beginning of magic state distillation is $T_{\mathrm{msd}} = 230 + 2d + 3d \approx 330$. 

\subsection{Comparison to linear in-place CCZ}
\begin{figure}
    \centering
    \includegraphics[width=\linewidth]{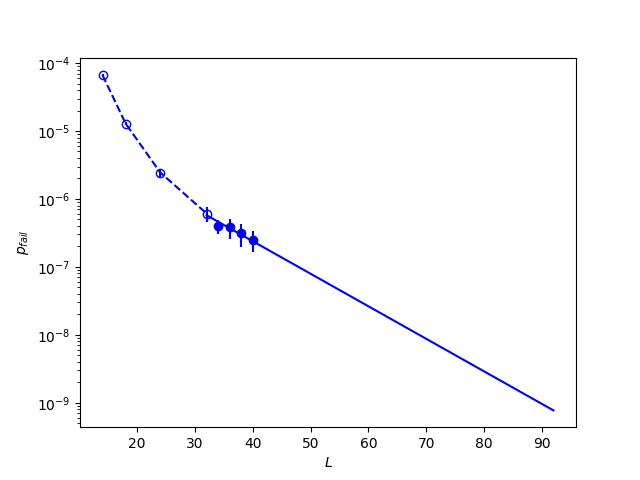}
    \caption{Logical $X$ error probability ($p_\textrm{fail}$) in a single surface code (the red code described above) vs distance ($L$) when performing a linear-time CCZ under phenomenological noise, using code from~\cite{scrubyNumericalImplementationJustTime2022,JITDecodingRepo} and physical error probability $p=4.67\times10^{-4}$. Between $1\times10^8$ and $2\times10^8$ samples were taken for each value of $L$. Empty points are data from \cite{scrubyNumericalImplementationJustTime2022} while filled points are new data produced for this work. The solid line is a linear fit to the five data points in the range $30 < L \leq 40$. We emphasise that this performance is dependent on the JIT decoder and not an inherent property of the linear-time CCZ, and so overheads would improve given an improved decoder. Data for this plot is available at \cite{DataRepo}.}
    \label{fig:jit_extrapolation}
\end{figure}

To estimate the code distance $d_{\textsc{ccz}}$ required to achieve a logical error rate $P_{\textsc{ccz}} \sim 3 \times 10^{-10}$ with the linear time CCZ, we can extrapolate from numerical data. The only prior data available is from Ref.~\cite{scrubyNumericalImplementationJustTime2022}. However, due to finite-size effects, meaningful extrapolation cannot be performed using only the distances simulated in that work. Therefore, we use the same code~\cite{JITDecodingRepo} to perform additional simulations at larger distances as shown in \cref{fig:jit_extrapolation}. Extrapolating from these data, we estimate the required distance to be $d_{\textsc{ccz}} \sim 100$. We emphasise that this estimate is unreliable for a number of reasons, such as the large amount of extrapolation employed and the lack of a circuit-level noise model in these simulations, and thus the real-world performance can be worse than this. Additionally, the slices simulated in Ref.~\cite{JITDecodingRepo} are not the same slices we have described above and so some amount of variation in performance can be expected to result from this. Regardless, we consider $d_{\textsc{ccz}} \sim 100$ to be a suitable lower bound on the code distance required and note that larger values of $d_{\textsc{ccz}}$ would not change any of the conclusions of the following analysis. 

Given $d_{\textsc{ccz}} \sim 100$, the footprint of each logical qubit during the linear CCZ gates is expanded into $d_{\textsc{ccz}} \times 2d_{\textsc{ccz}} \approx 5d \times 10d$. A natural architecture that allows the implementation of linear CCZ is shown in \cref{fig:ccz_layout}, where we have these $5d \times 10d$ patches laying side-by-side. Each patch in \cref{fig:ccz_layout} actually represents a stack of such patches using the looped pipeline architecture, CCZ gates can be implemented on almost all logical qubits in parallel (except for those at the boundary) by simultaneously sliding the layers in all patches. The same layer in different stacks will represent the same code out of the three and thus will slide in the same direction, while different layers in the same stack may slide in different directions since they can represent different code configurations out of the three. When not implementing the linear CCZ, the surface code patches will shrink to the size of $d \times d$ as shown in \cref{fig:log_layout}, where we have a corridor with width $4d$ in one direction and $9d$ in the other. 

As shown in \cref{fig:msd_layout}, this is actually enough to fit in one CCZ factory per 3 logical code patches, and in the looped pipeline architecture, this translates into a stack of CCZ factory per 3 stacks of logical code patches. If we want to perform CCZ gate over all qubits in this layout, we only need to output one magic state per magic state factory plus the time needed for gate teleportation, which is $T_{\mathrm{msd}} \approx 330$ as derived above. In \cref{sec:in_place_gate}, we have calculated the time overhead of the linear CCZ gates to be $T_{\mathrm{lin}} = 6d_{\textsc{ccz}} = 600$. We also need to add in the $d_{\textsc{ccz}}$ rounds needed for expanding the code patch from $d$ to $d_{\textsc{ccz}}$, thus the total time for the in-place CCZ is $T_{\mathrm{lin}}' = 7d_{\textsc{ccz}} = 700$. Hence, the space needed for implementing linear CCZ is enough for fitting a CCZ factory in place right next to the target qubits, and the time required for the CCZ factory to produce and teleport a CCZ gate is around $T_{\mathrm{lin}}'/T_{\mathrm{msd}} \approx 2$ times faster than linear CCZ.

\subsection{Comparison to linear CCZ factories}\label{sec:ccz_factory_comp}
\begin{figure}
    \centering
    \subfloat[Logical qubit layout \label{fig:log_layout}]{\includegraphics[width = 0.4\textwidth]{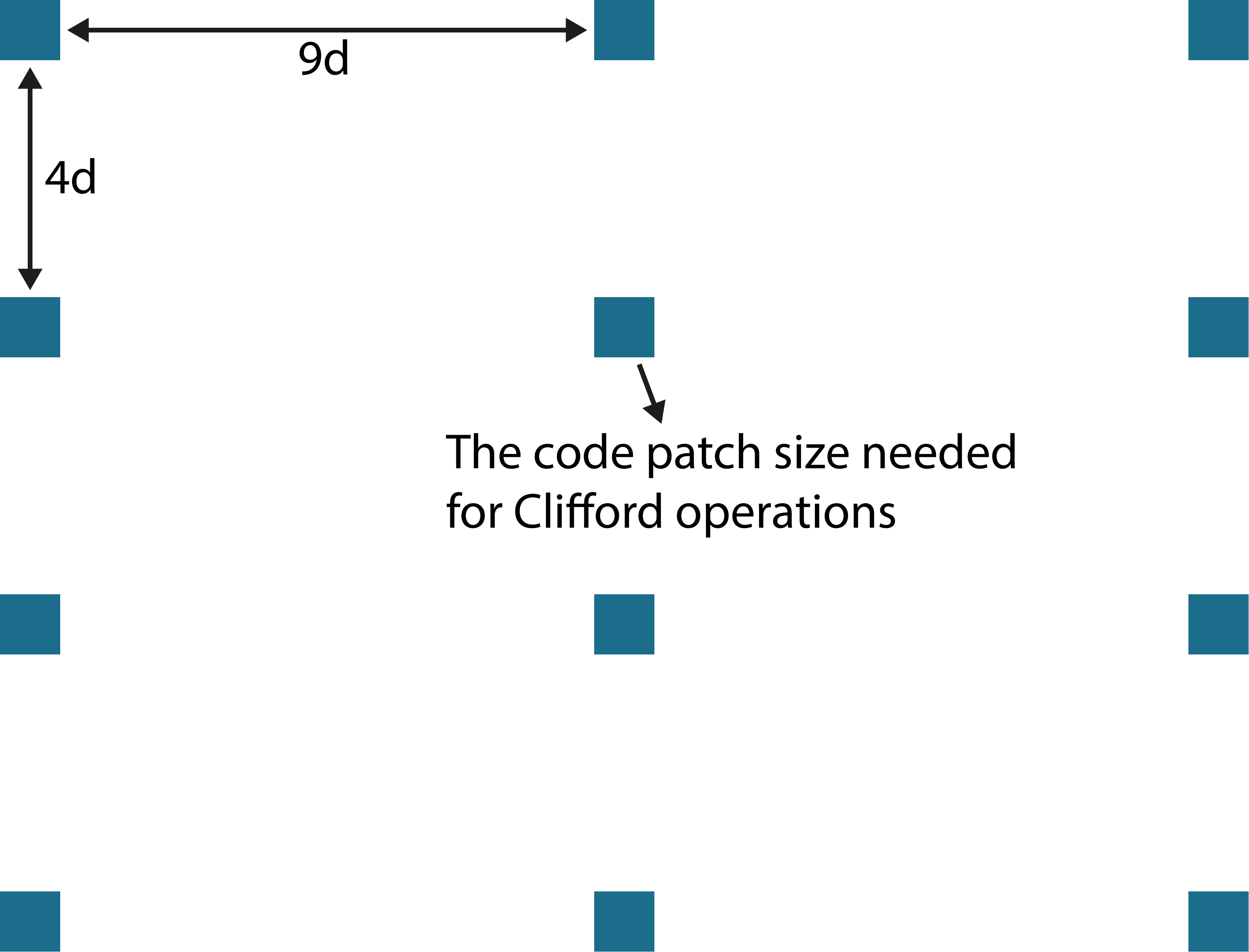}}\\
    \subfloat[Logical qubit expansion for linear CCZ \label{fig:ccz_layout}]{\includegraphics[width = 0.4\textwidth]{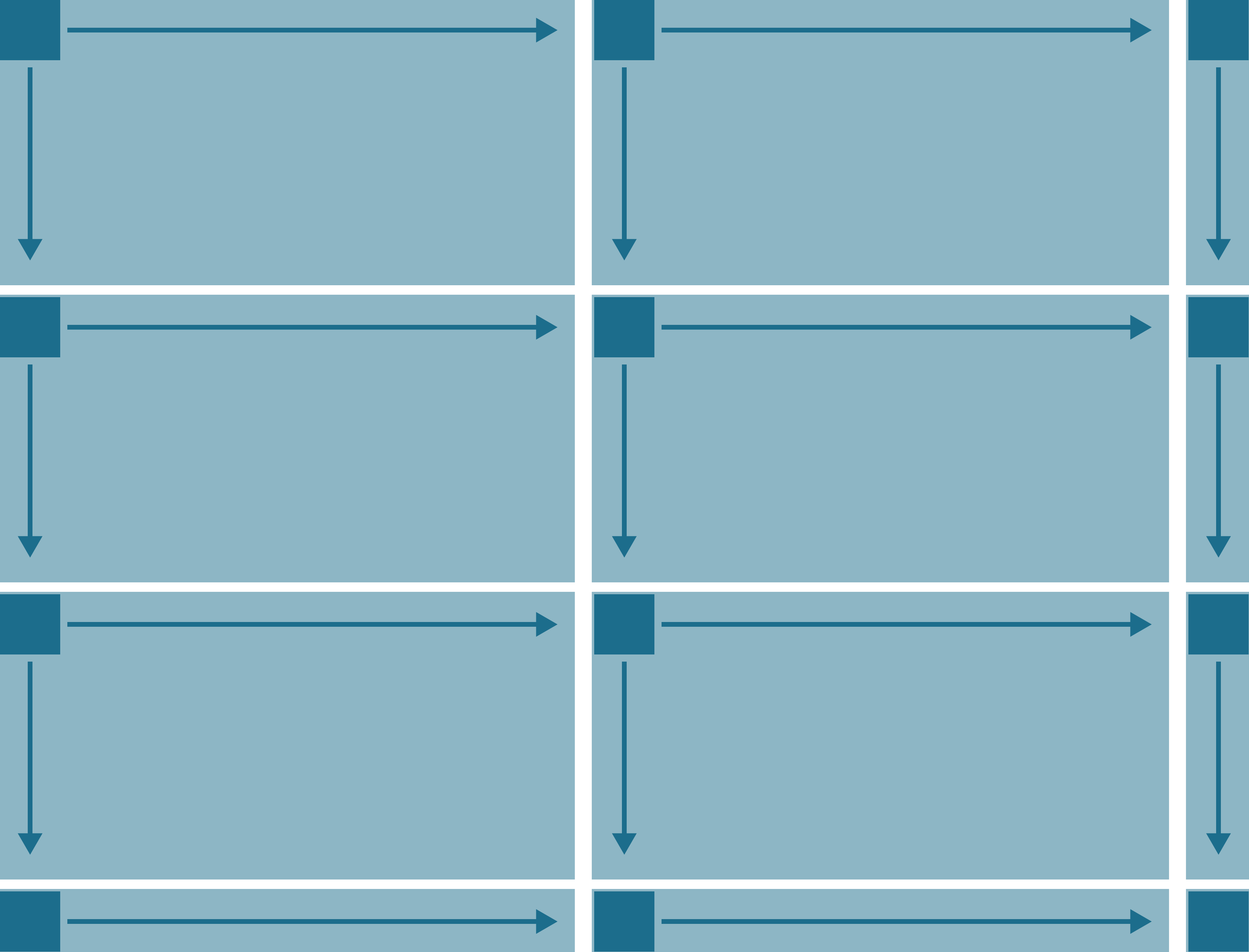}}\\
    \subfloat[Inserting magic state factories \label{fig:msd_layout}]{\includegraphics[width = 0.4\textwidth]{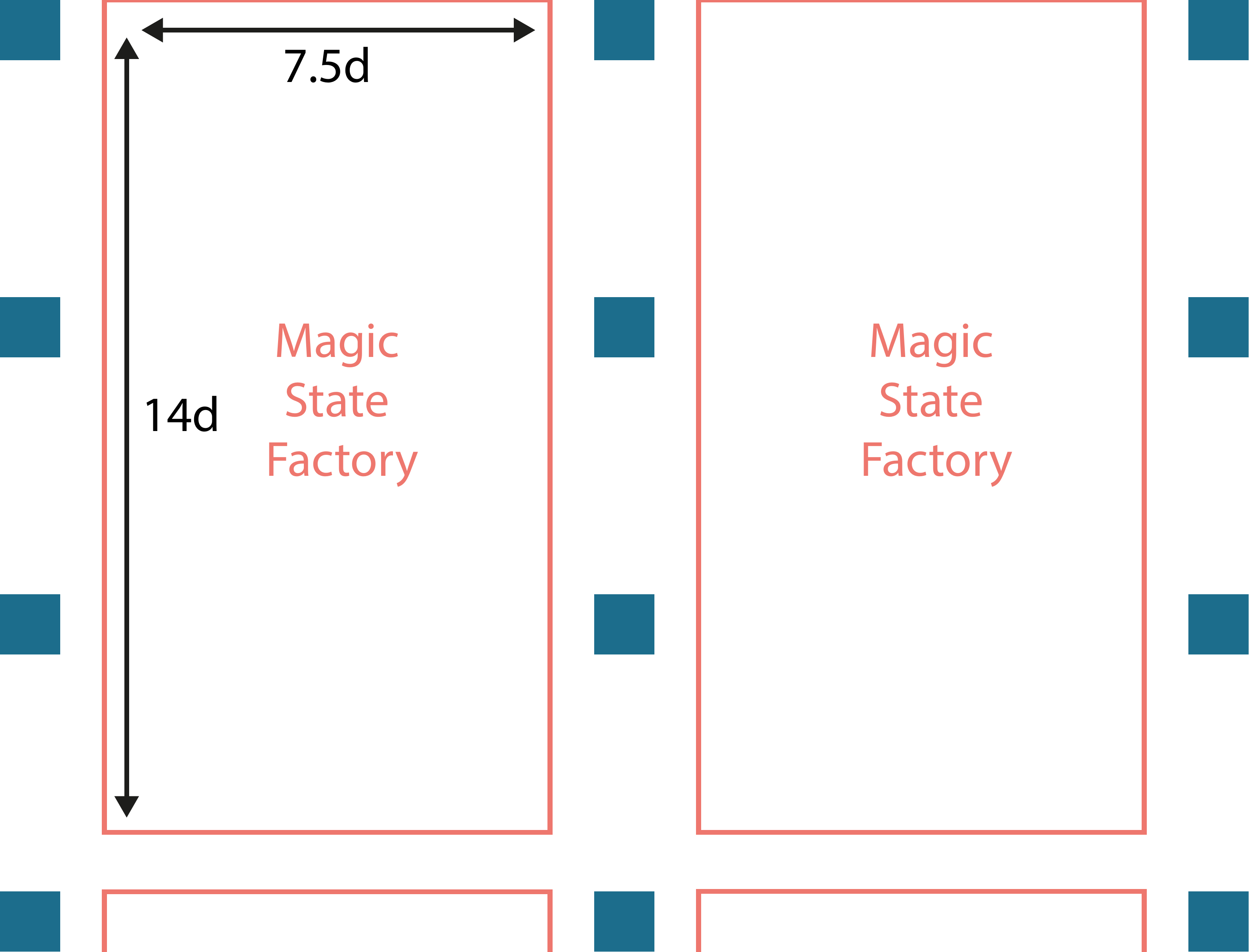}}
    \caption{Each patch here actually represents a stack of $K$ code layers where $K$ is the number of qubits per loop in the looped pipeline architecture. Similarly the magic state factory here actually contains a stack of $K$ magic state factories.}
    \label{fig:circuit}
\end{figure}

The $2$-times speed up here is for the case in which we want to apply parallel CCZ to all of the qubits. In practice, in most time steps we will not require so many CCZ gates such that one factory per three logical qubits is a gross overestimation. In fact, $14$ factories are enough for magic state distillation to be a non-rate-limiting step in our $6000$-qubit problem of factoring a 2048-bit integer using ripple carry adders ($28$ factories if using carry runways adders)~\cite{gidneyHowFactor20482021}. In these regimes, the architecture in \cref{fig:circuit} that allows in-place CCZ gates everywhere will lead to a massive waste of space. Instead, one may want to create space only at the logical qubits and at the time where we want to implement the CCZ gates, by moving other logical qubits around. 

Alternatively, we can also use linear CCZ to implement CCZ factories. We can construct such factories using a row in \cref{fig:ccz_layout}, a row of $M$ stacks with 3 layers per stack (and 6 layers during CCZ implementation) will give us $M-1$ linear CCZ factories. Hence, the average space overhead of a $\sim d_{\textsc{ccz}} \times 2Md_{\textsc{ccz}} / (M-1) \approx 50d^2$-size stack per linear CCZ factory (for $M \gg 1$). In this model of fitting three layers per stack, three distillation factories can be fit into a stack of the size $7.5d \times 14d$ as derived above, i.e. the space overhead is on average a $7.5d \times 14d/3 = 35 d^2$-size stack per CCZ distillation factory, this is 1.4 times smaller than the linear CCZ factory. Compared to the in-place CCZ gates, we do not need to expand the code patch from $d$ to $d_{\textsc{ccz}}$ in the linear CCZ factory, thus the time overhead is reduced from $T_{\mathrm{lin}}' = 7d_{\textsc{ccz}}$ to $T_{\mathrm{lin}} = 6d_{\textsc{ccz}}$. However, such a reduction is marginal, and thus the linear CCZ factory is still around $2$-times slower than the CCZ distillation factory. The routing overhead, etc, will be similar for both the linear CCZ and the CCZ distillation factories. In this setting, it is also possible to achieve a space-time tradeoff by using a standard 3DSC $X$ error decoder to check the correction produced by the JIT decoder after the gate is complete, and then post-select on outcomes where these two decoders agree\footnote{We would like to credit B. Brown for this insight.}. However, we do not expect that this can be used to make the linear CCZ competitive with distillation as the latter already outperforms the former in both space and time cost, and \cref{fig:jit_extrapolation} suggests that even a small reduction in code distance would lead to a significant reduction in the success probability. 

In our consideration here, we have made a range of favourable assumptions for the linear CCZ gates. 
We assume the expansion of layers into slices in the implementation of the linear CCZ does not exceed the number of qubits allowed per loop and does not increase the code cycle time. Furthermore, the distance $100$ that we are assuming here for linear CCZ is not based on a circuit-level error threshold, thus the actual distance needed for linear CCZ can be even higher. The CCZ factories we use here are simply the ones designed for a 2D layout, and we can fit multiple of them in a stack using the looped pipeline architecture. There is an additional possibility of taking advantage of the transversal entangling operations within the stack to further reduce the space overhead of the CCZ factories~\cite{caiLoopedPipelinesEnabling2023}, which we have not explored here.

\section{Conclusion}

We have shown that the linear-time CCZ gate of Ref.~\cite{brownFaulttolerantNoncliffordGate2020} is highly compatible with looped pipeline architectures, requiring a qubit shuttling schedule that is only slightly more complex than the standard surface code. This provides us with a valuable example of the utility of this architecture for the practical implementation of fault-tolerant CCZ, demonstrating how complex codes and fault-tolerant protocols can be straightforwardly implemented using a fairly minimal set of almost local operations. On a less positive note, the practical spacetime overheads for this particular fault-tolerant protocol still exceed those of magic-state distillation, even under a range of favourable assumptions. More specifically, if we want to perform in-place linear CCZ of error rate $P_{\textsc{ccz}} = 10^{-10}$ among all logical qubits, the distance needed for linear CCZ will create a corridor large enough to fit a CCZ distillation factory next to every triple of logical qubits as shown in \cref{fig:circuit}, so there is no advantage in the space overhead. The speed of CCZ distillation is two times faster than linear CCZ in our estimate.

The large overhead of the linear-time CCZ is mostly due to the low threshold and poor sub-threshold scaling of the JIT decoder. Hence, we expect the performance gap between these approaches to decrease when we have a smaller target CCZ error rate. For example, if we are targeting  $P_{\textsc{ccz}} = 10^{-7}$ instead, we will require $d_1 = 9$ and $d = 15$ for magic state distillation, which translates into a space overhead of $7d \times 14d$  and a time overhead of $(d_1 \times 2 + 1) \times 8.5 + 5d = 236.5$ code cycles. For linear CCZ, we will need $d_{\textsc{ccz}} = 50$, which translate into $3.3d \times 6.6d$ and $7d_{\textsc{ccz}}=350$ code cycles for space and time overhead, respectively. In this case, the space in the corridor created for implementing linear CCZ is no longer enough to fit one CCZ factory per $3$ logical qubits, thus we need to consider how CCZ resource states are stored, shared and routed when using CCZ factories. The time improvement is also reduced from $2$ times to $1.5$ times. The CCZ distillation factories approach will still have the advantage taking into full consideration the various additional factors mentioned at the end of \cref{section:resource}, but such an advantage is reduced compared to when $P_{\textsc{ccz}} = 10^{-10}$. 

One also needs to keep in mind that magic state distillation, being a method that uses post-selection, an exponential suppression in the output error rate will require an exponential increase in the number of trials (i.e. an exponential decrease in the pass rate of the checks). This characteristic is even more prominent in the recent related cultivation scheme~\cite{gidneyMagicStateCultivation2024}, which only targets a specific range of output error rates. On the other hand, the linear CCZ being a fault-tolerant approach only requires a linear increase in the code distance to achieve exponentially stronger error suppression, even though such an advantage over magic state distillation will only occur at ultra-small target CCZ error rate far below the regime of practical interests with the current schemes. 
Thus more interestingly, one can look into ways to merge the philosophy of post-selection from the distillation/cultivation schemes into the linear CCZ schemes in order to achieve a better balance between asymptotically good behaviour and practical performance. The decoder cross-checking method mentioned in \cref{sec:ccz_factory_comp} is one such idea.

It is also worth noting that the JIT decoder used to obtain these performance estimates is (to our knowledge) the only such decoder that has been investigated numerically, and there is no reason to believe that it is optimal, or even close to optimal. Improved decoders that have higher thresholds or better sub-threshold scaling would also lead to significant reductions in the associated resource costs for the linear-time CCZ. It seems unlikely that such improvements could by themselves make the linear-time gate competitive with magic state distillation in the practical regime, especially with the continuous progress in magic state distillation~\cite{gidneyMagicStateCultivation2024}. However, when combined with other factors (e.g. lower target logical error rates as discussed above), there may exist regimes where these approaches have a similar cost.

\section*{Acknowledgements}
The authors would like to thank Sam Jaques, Mike Vasmer and Ben Brown for insightful discussions and suggestions.
ZC acknowledges support from the EPSRC QCS Hub EP/T001062/1, EPSRC projects Robust and Reliable Quantum Computing (RoaRQ, EP/W032635/1), Software Enabling Early Quantum Advantage (SEEQA, EP/Y004655/1) and the Junior Research Fellowship from St John’s College, Oxford. TRS acknowledges support from the JST Moonshot
R\&D Grant [grant number JPMJMS2061]. KN acknowledges support from JSPS KAKENHI [grant number
21H04880]. Numerical results presented in this work were partially obtained using the HPC resources provided by the Scientific Computing and Data Analysis section of the Research Support Division at OIST.

\bibliography{ref}
\end{document}

%% file: JIT_background.tex
The linear-time CCZ is best understood as a reordering of the operations involved in the constant-time (transversal) CCZ between three copies of the 3D surface code~\cite{vasmerThreedimensionalSurfaceCodes2019}. A procedure that prepares a CCZ magic state using the constant-time gate consists of the following steps:

\begin{itemize}
    \item Initialise all physical qubits in $\ket{+}$, measure all $Z$ stabilisers and apply an $X$ correction based on the outcomes of these measurements. This will prepare all three 3D codes in the logical $\ket{+}$ state. 
    \item Apply transversal physical CCZ between the three 3D codes 
    \item Measure out some of the physical qubits to project the states of the 3D codes to states of 2D codes supported on their boundaries. Alternatively, the three 3D codes can be entangled with three 2D codes and then all the qubits of the 3D codes can be measured out to teleport their states to the 2D codes. 
\end{itemize}

In the linear-time gate, the sequence of operations experienced by each qubit is (in principle\footnote{In practice the different decoding strategies used in the two cases will sometimes result in inequivalent corrections}) identical. The difference is that, for each of the three codes, the qubits are not all initialised at once. Instead, a thin slice of the 3D code is initialised, the physical CCZs are performed, and the physical qubits on the ``bottom'' of the slice are measured out while a new layer of qubits is initialised at the ``top'' (we will use the term ``layer'' to refer to a strictly 2D code while ``slice'' refers to a 3D code composed of a small number of layers). Individual CCZ gates are only applied to qubits after all $Z$ stabilisers supported on that qubit have been measured, so the order of operations for each individual qubit is identical to the 3D case. After $O(d)$ timesteps, the spacetime history of the code will look like the full-distance 3D code and the final layer of qubits to be initialised will end up in the same state as the final 2D code in the constant-time procedure. 

\begin{figure}
    \centering
    \includegraphics[width=.35\textwidth]{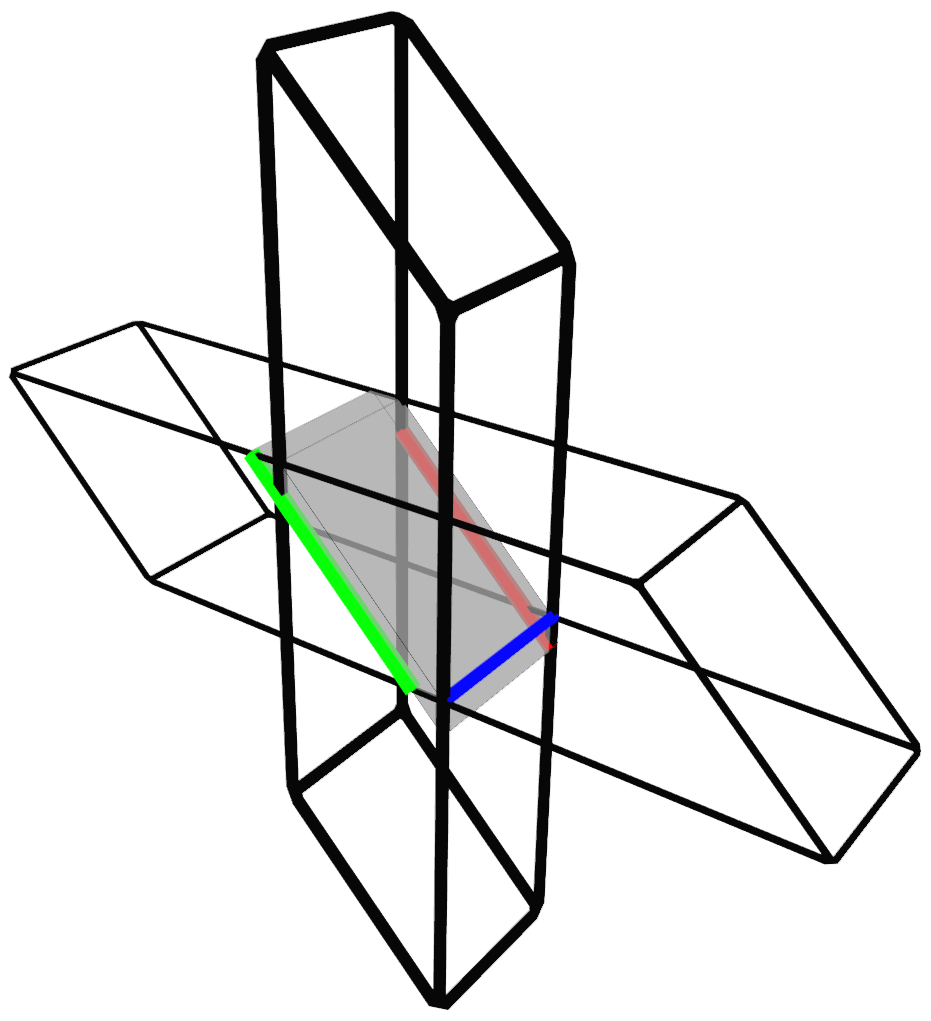}
    \caption{An example of three 3D surface code volumes which support a transversal CCZ. The thick black lines represent two overlapping codes while the third code is oriented in an orthogonal direction. When interpreted as a spacetime diagram for three 2D codes time for the two vertical codes runs bottom-to-top while time for the horizontal code runs left to right. A slice through the three codes is shown in grey and three logical $Z$ operators are shown as coloured lines (note that although the red and green operators are parallel they run between different pairs of boundaries).}
    \label{fig:volumes}
\end{figure}

A practical implementation of this procedure would not be quite so simple. One notable issue comes from the fact that for the procedure to be fault-tolerant our slices through each code must contain distance $d$ representations of the logical operators of that code. In each of these codes, the logical $Z$ operators are stringlike, so we require there to be a $Z$ logical operator representation lying in the plane normal to the time direction. However, for CCZ between three 3D surface codes to be transversal, the logical $Z$ operators of the three codes must all be perpendicular and so we cannot choose a single time direction that preserves code distance in all three codes. Instead, at least one of the three codes must have a different time direction to the other two and our slices should be taken normal to the net time direction (see \cref{fig:volumes}). We must additionally extend each of the three 3D codes along their respective time directions in order to preserve distance throughout the procedure and the result is that the three codes overlap on only a subset of qubits and CCZ is applied only between the qubits in this subset. Qubits not in this subset exist only to preserve the code distance and are not strictly necessary for the logical operation, meaning that the spacetime cost of the linear-time procedure is higher than for the constant-time version. Altogether this gives a procedure in which we begin with three patches of 2D surface code (two overlapping and one disjoint) which move towards, through each other, and then apart again as the gate is performed. The point of maximum overlap between the three patches is shown by the grey slice in \cref{fig:volumes}.

In \cref{section:architecture} we describe in detail how pipelining can allow for the implementation of a valid set of 3D slices in a strictly 2D device with local connections. For more details on the decoding strategies and fault-tolerance of the procedure readers should consult~\cite{brownFaulttolerantNoncliffordGate2020,scrubyNumericalImplementationJustTime2022}. 

%% file: Looped_pipeline.tex
\begin{figure}[htbp]
    \centering
    \subfloat[A five-qubit array in linear shuttling tracks. \label{fig:linear_shuttle}]{\includegraphics[width = 0.45\textwidth]{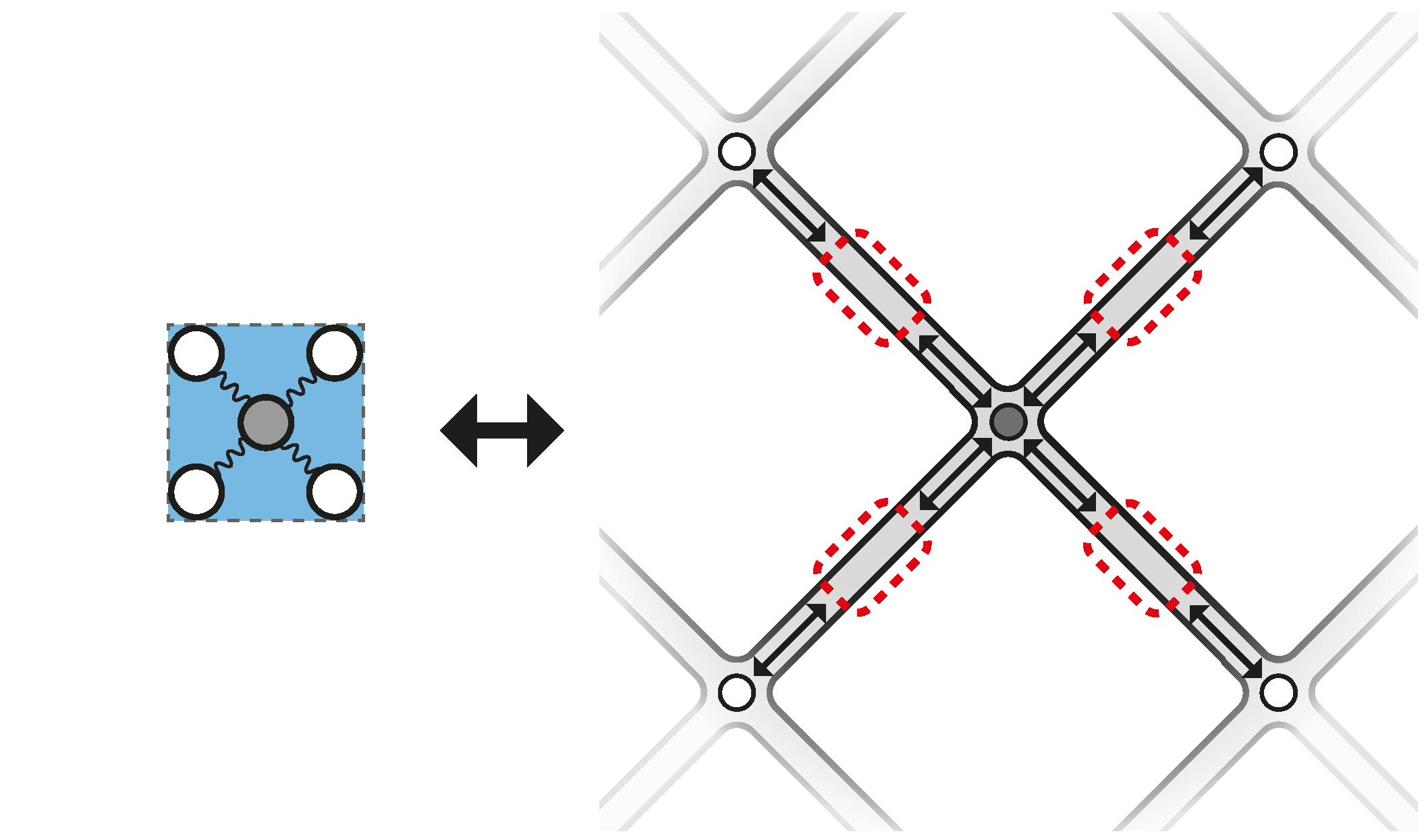}}\\
    \subfloat[A five-qubit array in shuttling loops. \label{fig:loop_shuttle}]{\includegraphics[width = 0.45\textwidth]{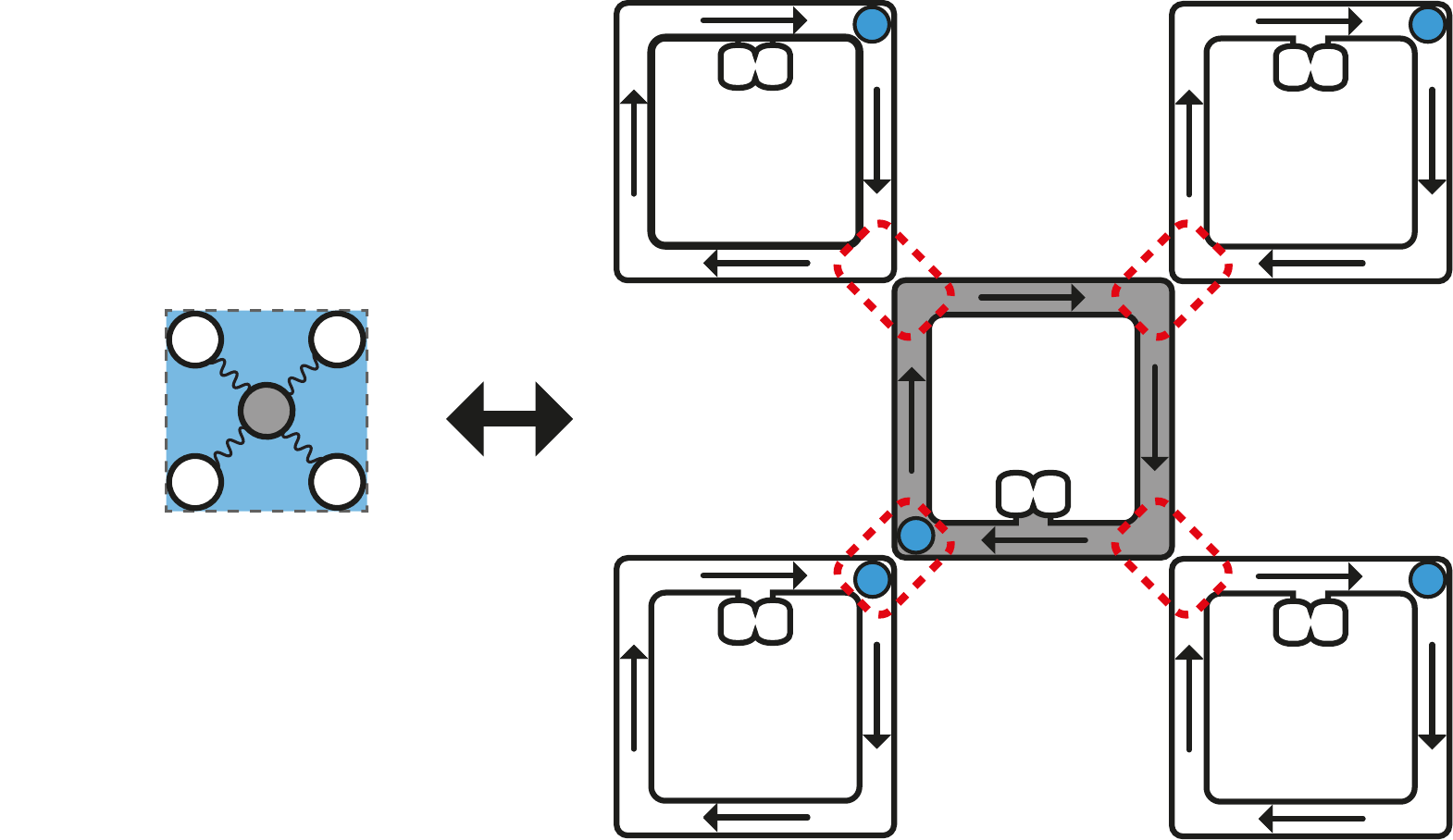}}\\
    \subfloat[Four five-qubit arrays in shuttling loops using pipelining .\label{fig:loop_pipeline}]{\includegraphics[width = 0.45\textwidth]{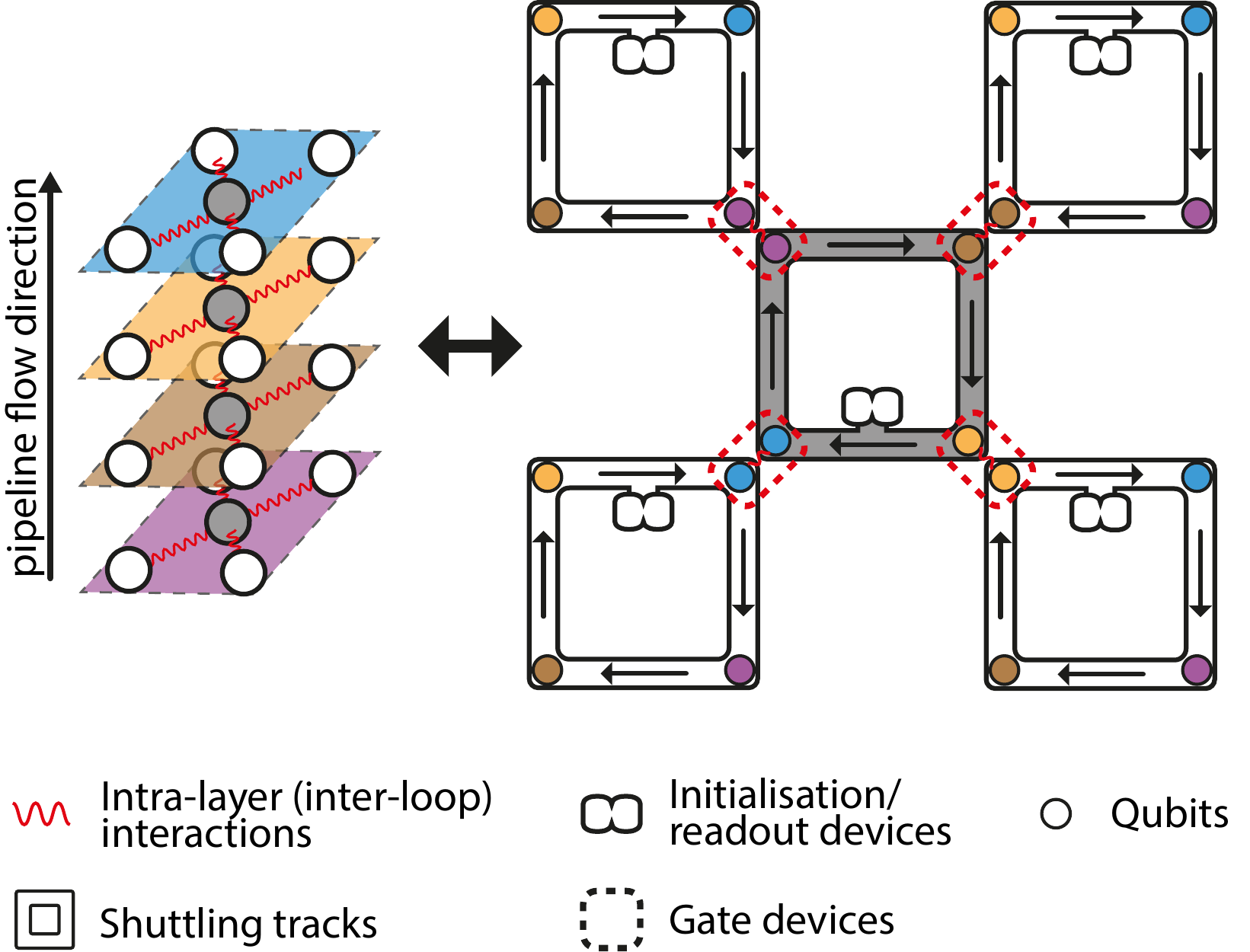}}
    \caption{Five-qubit arrays implemented using different shuttling architectures. Multiple qubits can be stored on the same loop through pipelining, which corresponds to multiple layers of five-qubit arrays.}
    \label{fig:array_pipeline_both}
\end{figure}

In many hardware architectures like semi-conductor spin qubits and trapped-ion qubits, in order to provide enough spaces for the control electronics and to avoid cross-talks, the qubits are often spaced out using linear shuttling highways and only brought together for two-qubits interactions~\cite{buonacorsiNetworkArchitectureTopological2019,boterSpiderwebArraySparse2022,kielpinskiArchitectureLargescaleIontrap2002,lekitschBlueprintMicrowaveTrapped2017,kaushalShuttlingbasedTrappedionQuantum2020,ryan-andersonRealizationRealTimeFaultTolerant2021,hilderFaultTolerantParityReadout2022} as shown in \cref{fig:linear_shuttle}. It was proposed in \cite{caiLoopedPipelinesEnabling2023} that the above challenges also be solved by having qubits running around shuttling loops in synchronisation as shown in \cref{fig:loop_shuttle}. As the qubit at the centre loop moves around, it will meet and interact with qubits in the neighbouring loops one by one, carrying out all possible nearest-neighbour interactions after moving one full cycle around the loop. At first glance, such an architecture based on shuttling loops is no different from the linear version since they use the same amount of hardware. However, using the exact same looped architecture for one qubit array, we can fit additional qubits into each loop as shown in \cref{fig:loop_pipeline}. There the yellow qubits follow the exact same path as the blue qubits, just one step behind, thus forming a yellow-qubit array that is entirely independent of the blue array we had. In a similar way, as we fit $K$ qubits in each loop, we can now store and process a stack of $K$ qubit arrays in parallel with the exact same connectivity. Such a looped pipelined architecture thus allows us to increase the qubit density by $K$ times without needing to add any additional hardware. 

\begin{figure}[htbp]
    \centering
    \includegraphics[width = 0.45\textwidth]{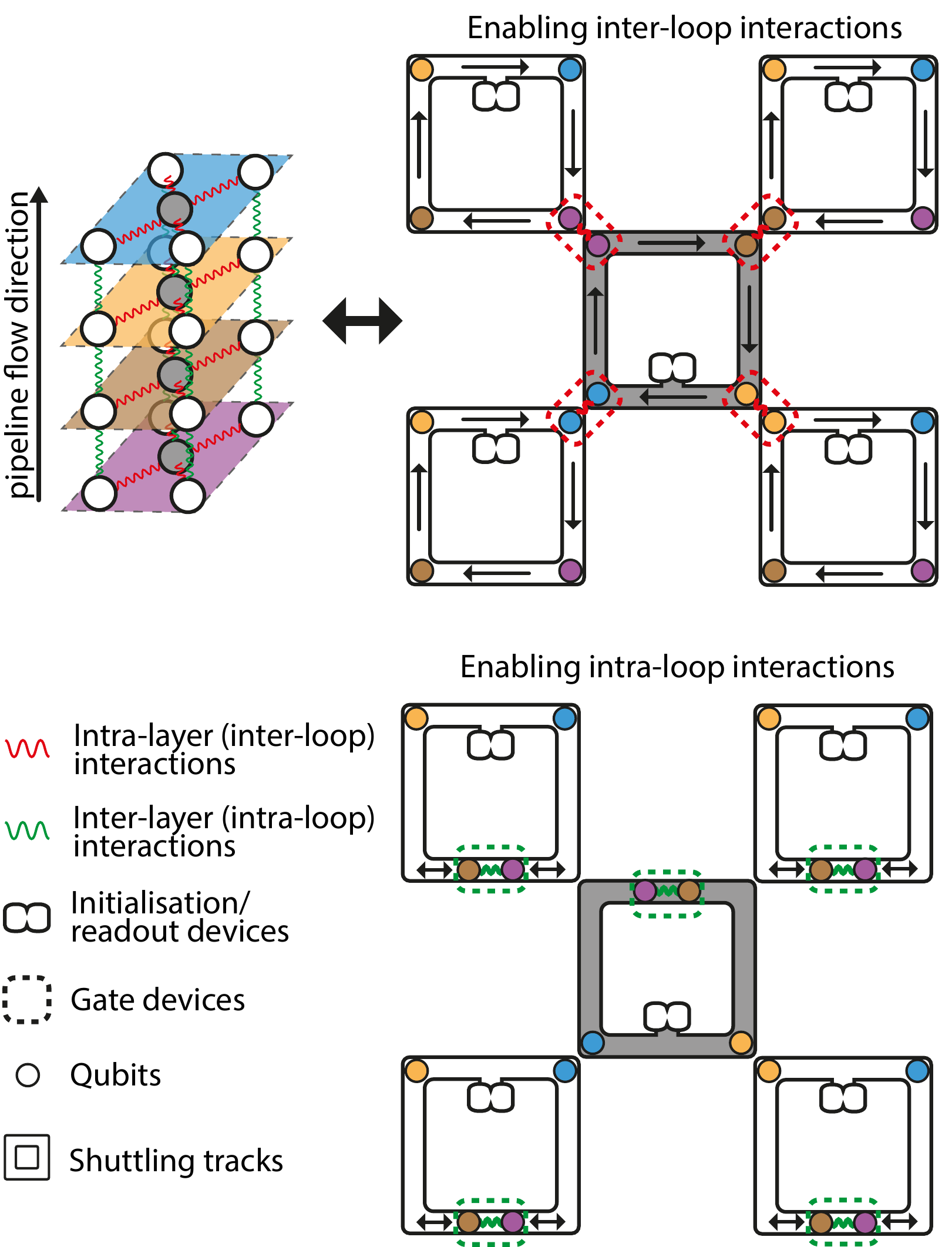}
    \caption{Interactions within layers are enabled by inter-loop (red) interactions while transversal interaction in between layers are enabled by intra-loop (green) interactions.
    }
    \label{fig:array_pipeline_intra}
\end{figure}

Now we have seen that interacting the corresponding qubits between different loops (inter-loop interactions) allows for intra-layer interactions within the stack of qubit arrays. If we interact with the qubits within the same loops (intra-loop interactions) as shown in \cref{fig:array_pipeline_intra}, then we are effectively interacting with the corresponding qubits in between the different layers of qubits, i.e. performing transversal operations between qubits. This effectively gives us 3D qubit connectivity on a 2D platform through pipelining. 

One needs to note that even though shuttling might be useful for arbitrarily increasing qubit connectivity in small experiments, it remains a challenge to devise efficient shuttling schemes in a scalable way taking into account the shuttled distance, scheduling, etc. The looped pipeline architecture is such a \emph{scalable} scheme for increasing qubit connectivity that is applicable to platforms like silicon spins, trapped ions and neutral atoms.

%% file: architecture.tex
In this section we describe how to use the looped pipeline architecture described in the previous section to implement the error correcting codes and linear-time non-Clifford gate described in~\cite{brownFaulttolerantNoncliffordGate2020}. 

\subsection{Code Details}

\begin{figure}
    \centering
    \includegraphics[width=.2\textwidth]{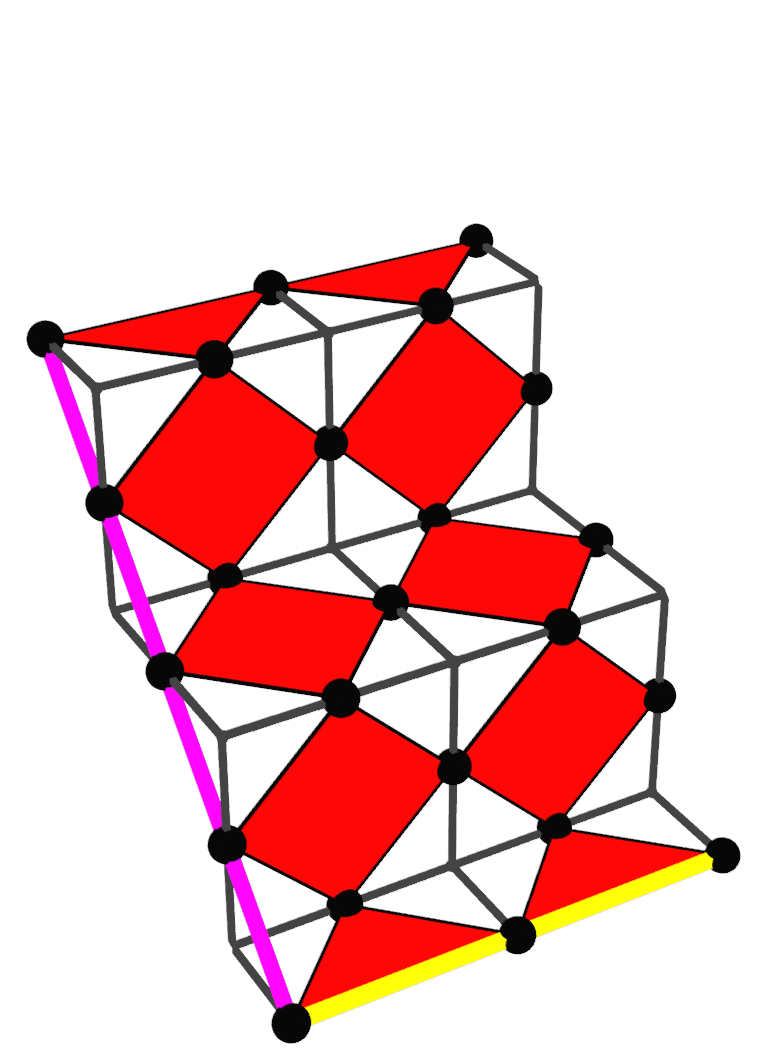}
    ~
    \includegraphics[width=.2\textwidth]{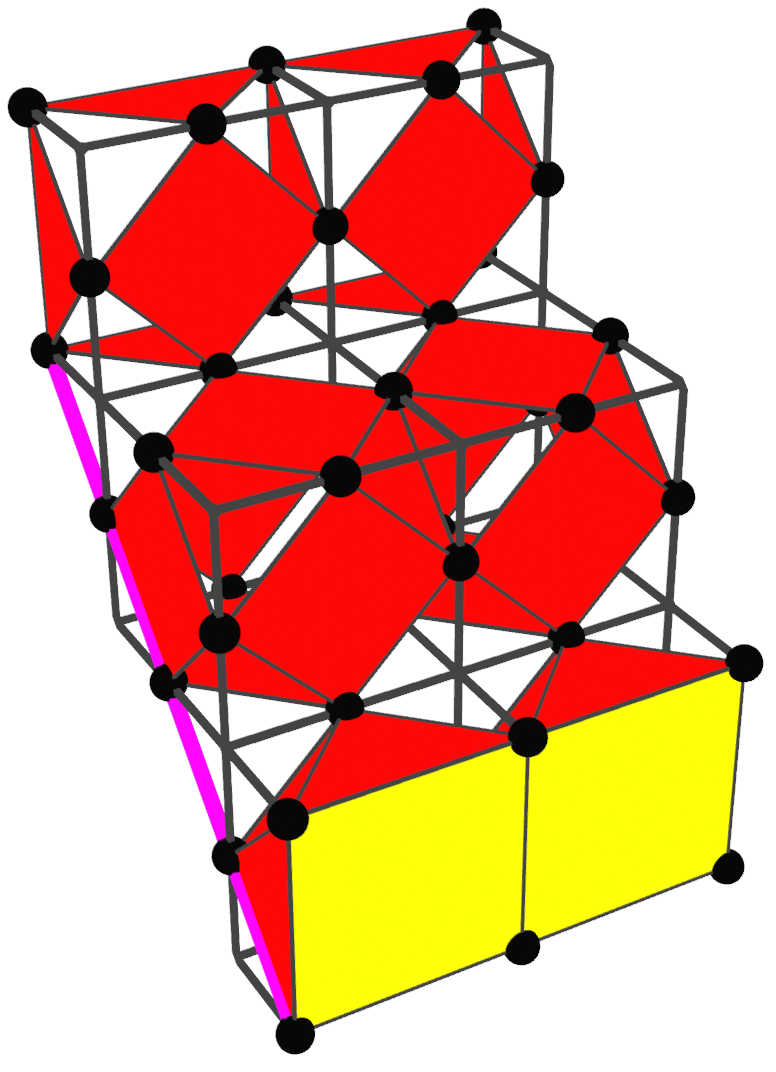}
    \\
    \vspace{5mm}
    \includegraphics[width=.2\textwidth]{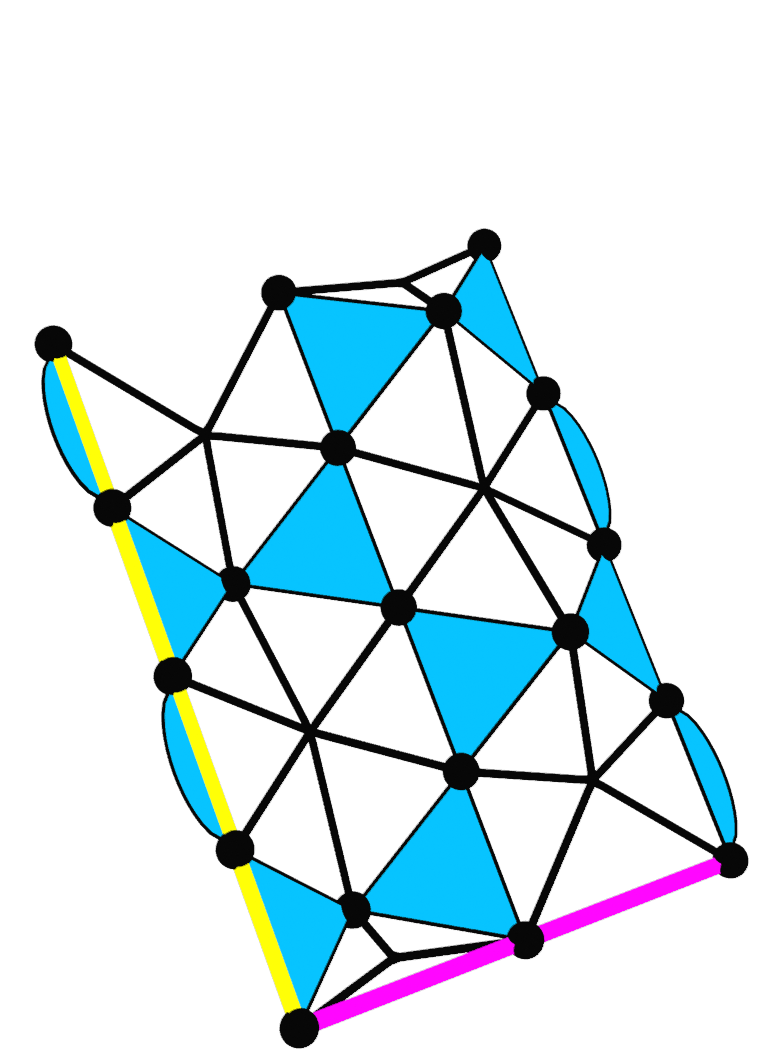}
    ~
    \includegraphics[width=.2\textwidth]{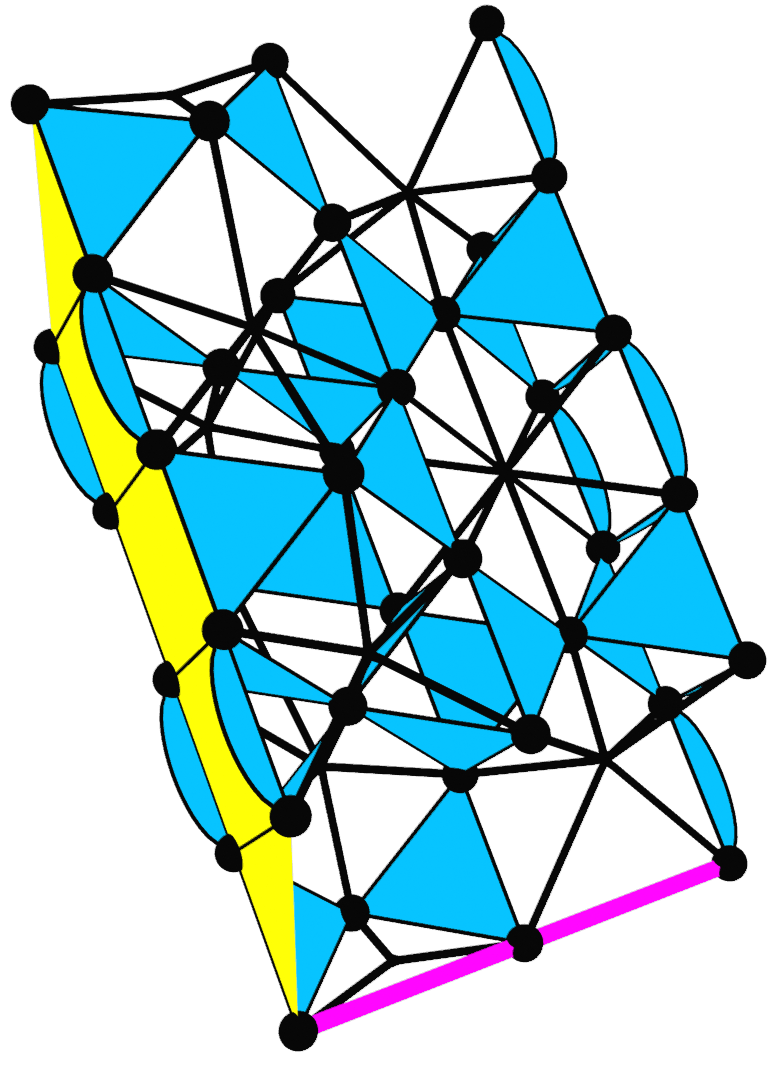}
    \\
    \vspace{5mm}
    \includegraphics[width=.2\textwidth]{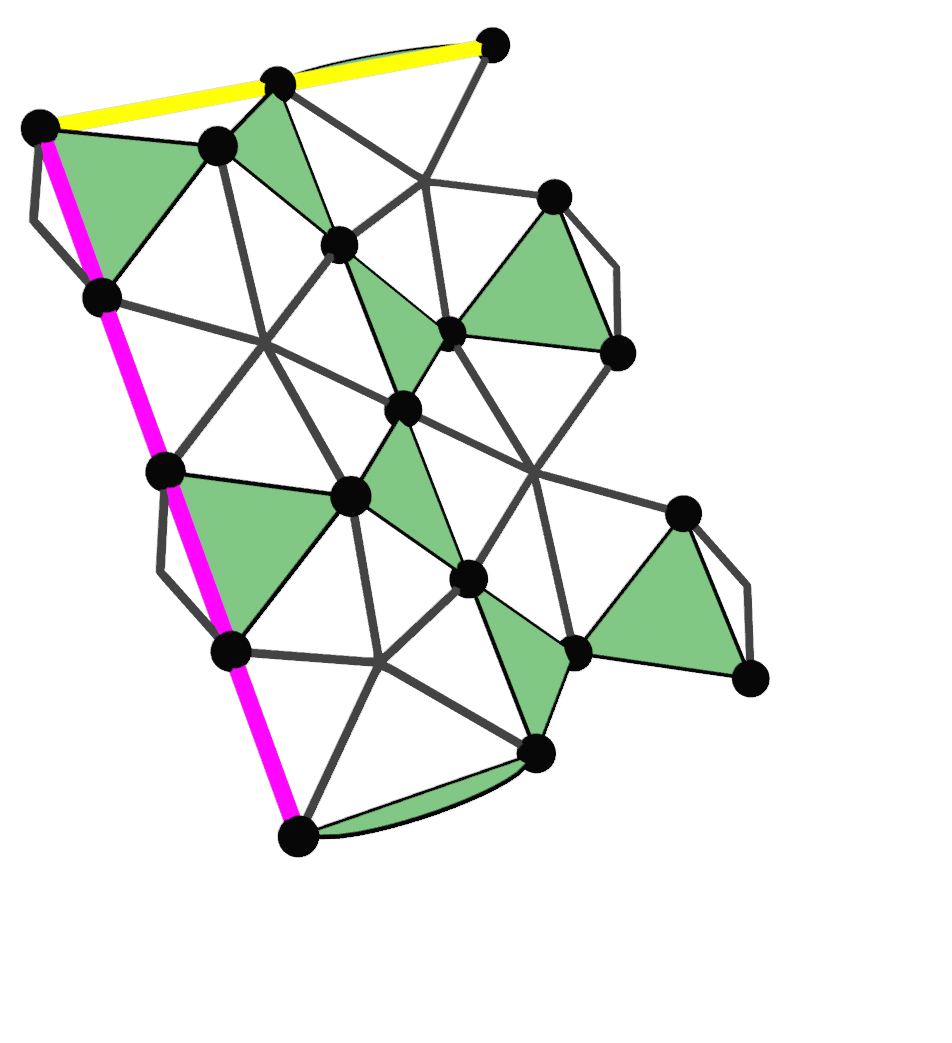}
    ~
    \includegraphics[width=.2\textwidth]{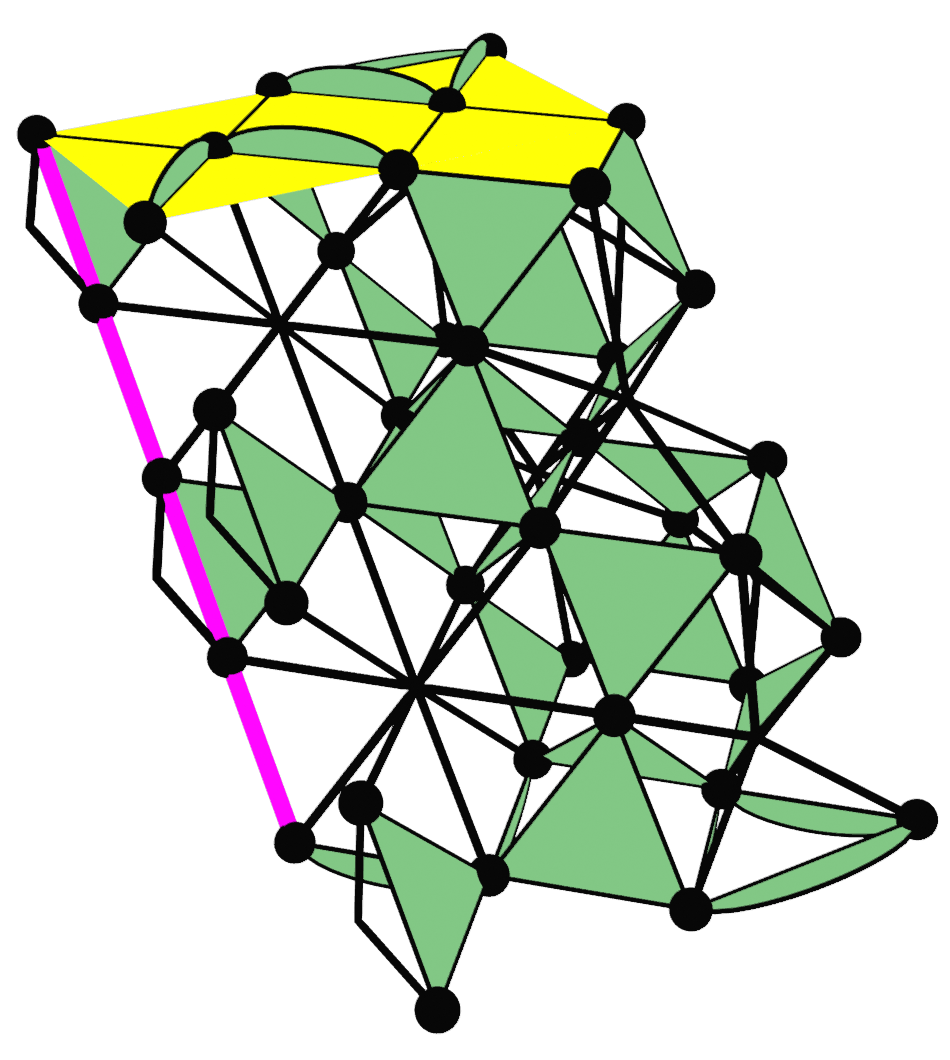}
    \\
    \vspace{5mm}
    \includegraphics[width=.2\textwidth]{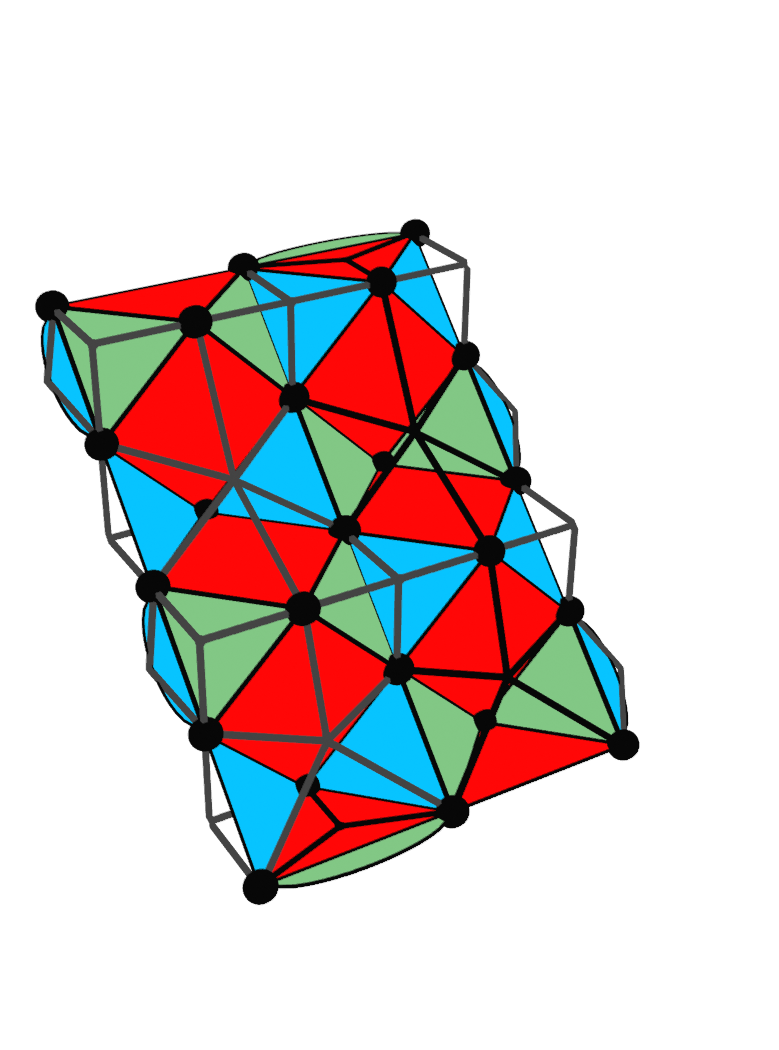}
    ~
    \includegraphics[width=.2\textwidth]{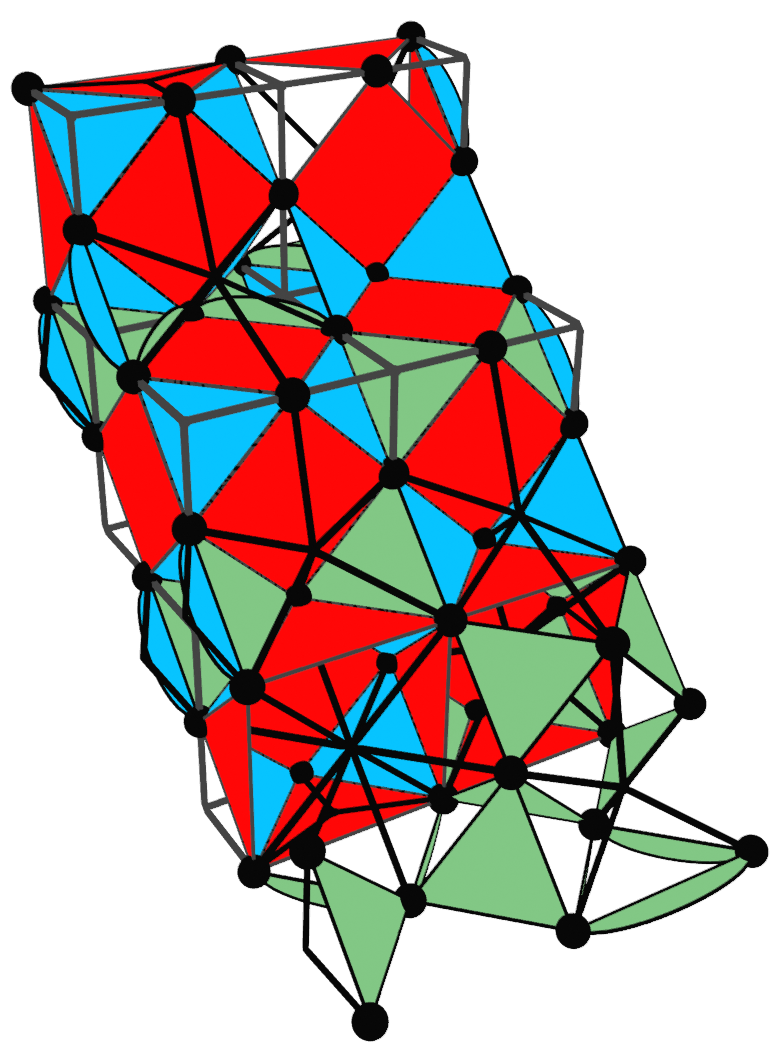}
    \caption{Layers and slices through each of the three codes + their overlap. Black circles are qubits, black stars are $X$ stabilisers and coloured faces are $Z$ stabilisers (different colours are only used to differentiate the three codes). Logical $Z$ representations are shown by pink lines and logical $X$ representations are shown by yellow lines/membranes.}
    \label{fig:slices}
\end{figure}

\begin{figure}
    \centering
    \includegraphics[width=.48\textwidth]{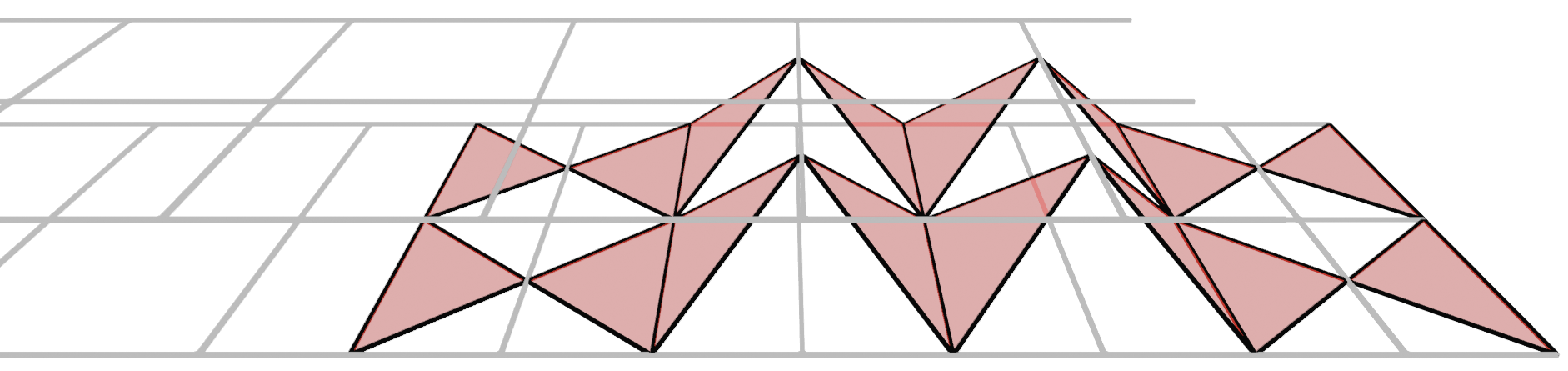}
    \\
    \vspace{5mm}
    \includegraphics[width=.48\textwidth]{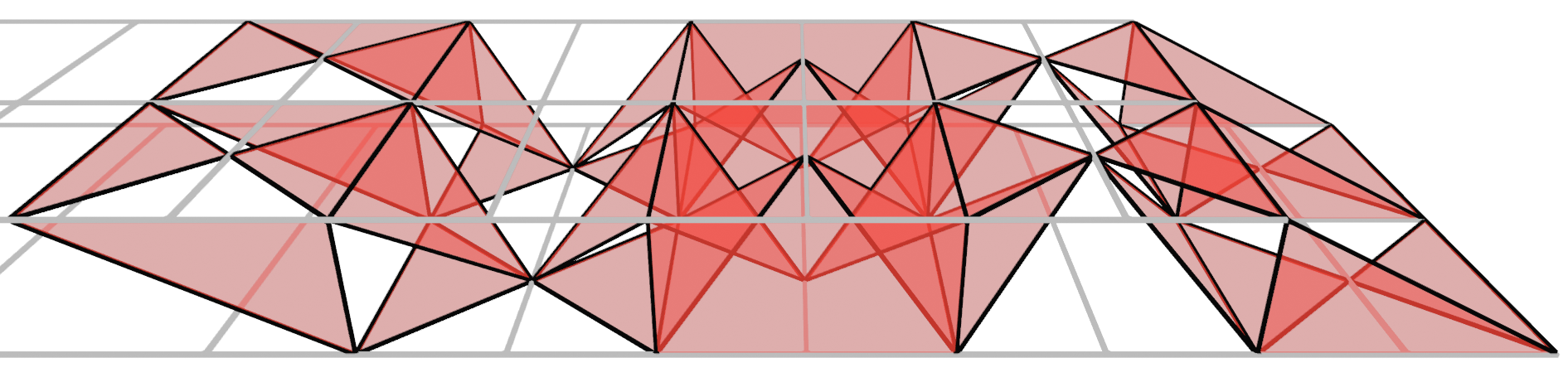}
    \\
    \vspace{5mm}
    \includegraphics[width=.48\textwidth]{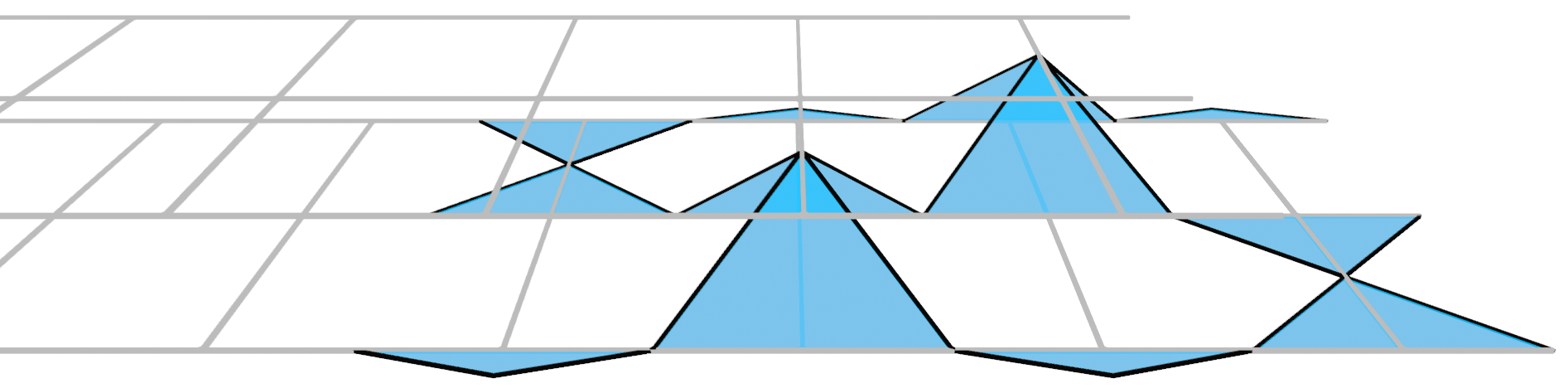}
    \\
    \vspace{5mm}
    \includegraphics[width=.48\textwidth]{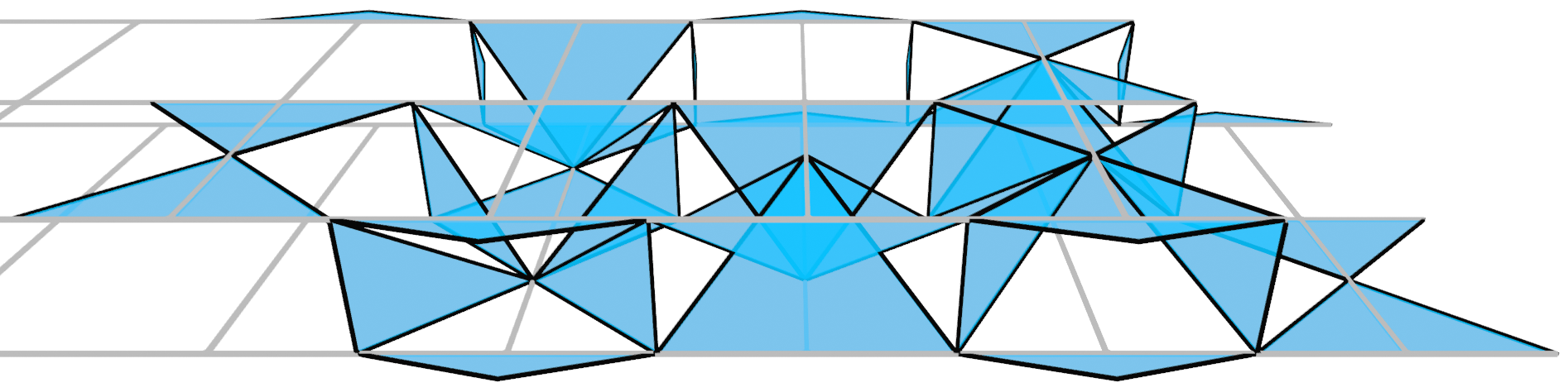}
    \\
    \vspace{5mm}
    \includegraphics[width=.48\textwidth]{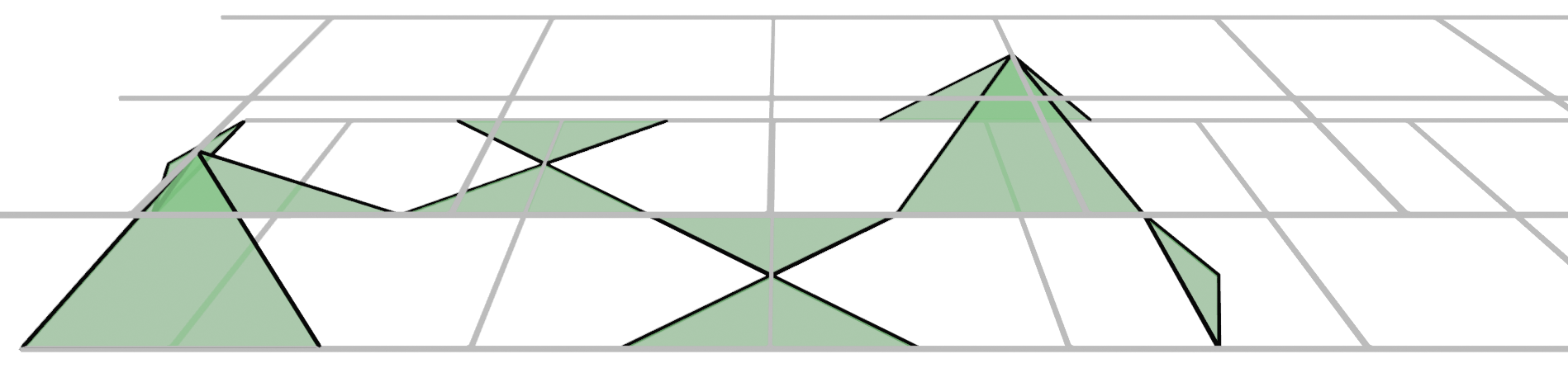}
    \\
    \vspace{5mm}
    \includegraphics[width=.48\textwidth]{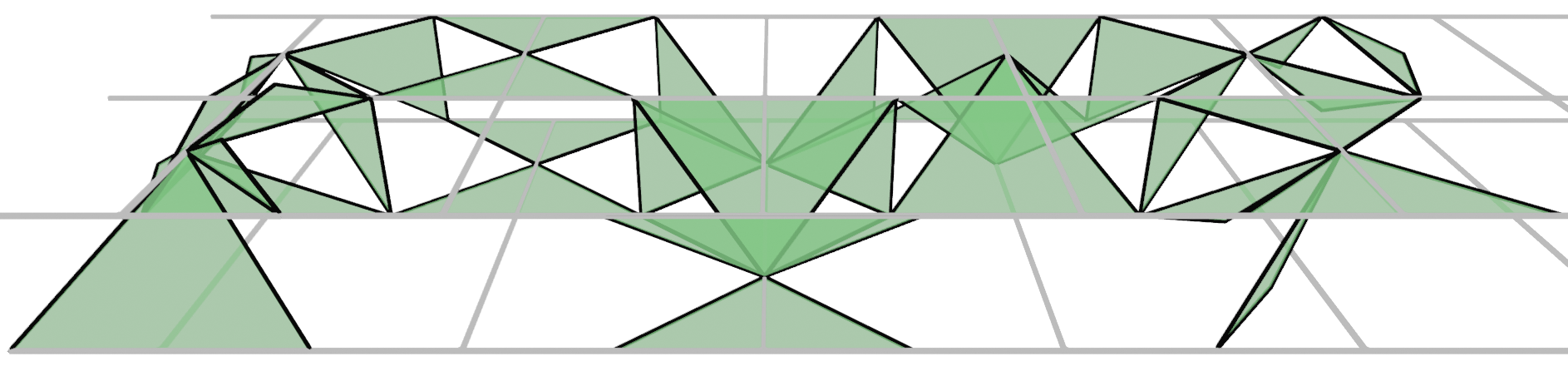}
    \\
    \vspace{5mm}
    \includegraphics[width=.48\textwidth]{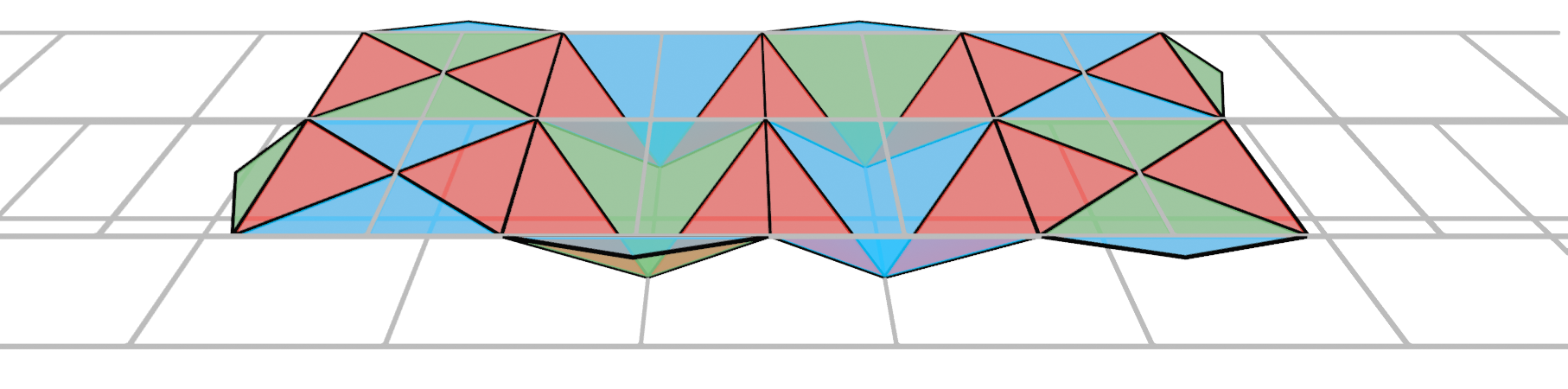}
    \\
    \vspace{5mm}
    \includegraphics[width=.48\textwidth]{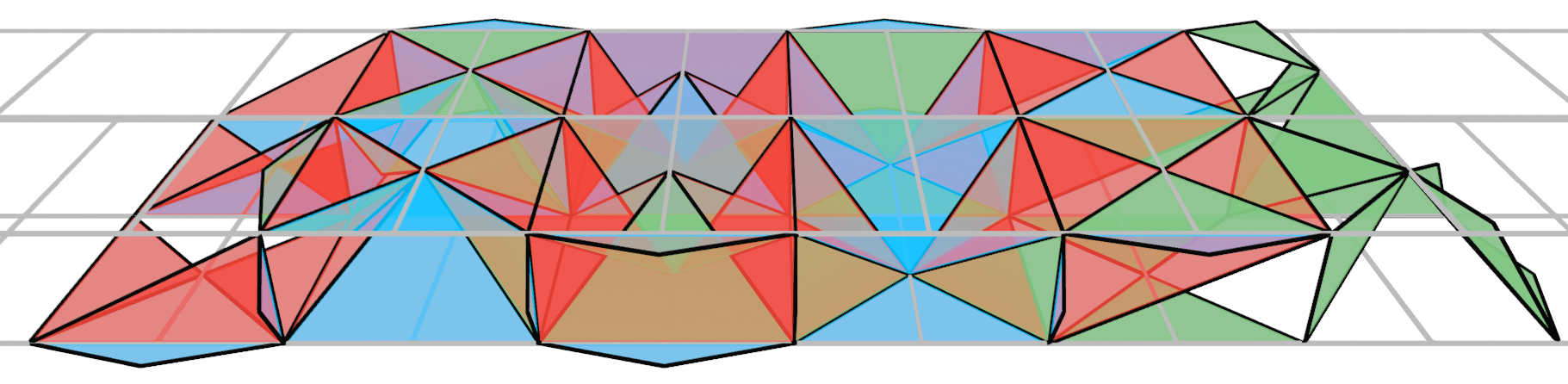}
    \caption{Mapping of the layers and slices from \cref{fig:slices} to a pair of square grids with qubits on edges. Only $Z$ stabilisers are shown. The assignment of the qubits of the first layer to the top or bottom grid is arbitrary, but the choice shown here gives a nice symmetry between this and subsequent layers.}
    \label{fig:grids}
\end{figure}

The slices of 3D code used to perform the linear-time CCZ cannot be chosen arbitrarily and only certain choices will preserve the code distance while also being compatible with the CCZ gate and JIT decoder. Two proposals for valid slices exist in the literature~\cite{brownFaulttolerantNoncliffordGate2020,scrubyNumericalImplementationJustTime2022}. The first (one that is consistent with \cref{fig:volumes}) has the advantage of being composed of only two layers per slice, whereas the second requires three layers of qubits per slice but allows for a qubit initialisation and measurement sequence that is synchronised across the three codes. In this work, we will use the former set of slices as fewer layers lead to fewer qubits per loop, which should simplify scheduling and reduce errors. 

Examples of distance-3 single layers and two-layer slices for the three codes are shown in \cref{fig:slices}, which have different lattice structures~\cite{vasmerThreedimensionalSurfaceCodes2019}. In the first of the three (red faces) the 2D layers are equivalent to the standard 2D surface code with qubits on edges, $Z$ stabilisers on faces and $X$ stabilisers on vertices. The 2D layers in the other two codes (blue and green faces) are defined with qubits on the vertices of a kagome lattice, $Z$ stabilisers on triangular faces and $X$ stabilisers on hexagonal faces. These two codes are mirrored with respect to each other. 

We assign a vertical time direction to the red and blue codes and a horizontal time direction to the green code. In each case the 3D slice is obtained by taking a copy of the layer and translating it one lattice unit in the relevant time direction (and adding any appropriate intermediate stabilisers from the 3D code bulk). For the blue and green codes we must also mirror the translated layer along one axis and add extra qubits which lie between the top and bottom layers but are not part of either layer. Boundary stabilisers for each of the three codes are chosen so that logical $Z$ operators run between a different pair of boundaries in each case (more obvious in the slices than in the layers). The final two images of \cref{fig:slices} show the layers and slices at their point of maximal overlap (corresponding to the shaded slice position shown in \cref{fig:volumes}). 

\subsection{Pipelined Implementation}

Before describing an implementation of these slices using a pipelined architecture it is helpful to first consider a mapping of to a pair of square grids with qubits on edges as shown in \cref{fig:grids}. The first red and blue layers are initialised at the right-hand end of the two grids and subsequent layers are shifted to the left relative to these first layers. The first layer of the green code is initialised at the left-hand end of the grids and subsequent layers will be shifted to the right. This motion corresponds to the projection of the layers and slices onto the plane normal to the sum of both time direction vectors in \cref{fig:volumes}.

To obtain a pipelined implementation of these layers and slices, we first combine the two grids from \cref{fig:grids} into a single grid (by associating edges which differ only in their $z$ coordinate). Then we assign one loop to each vertex, edge and face of this grid. Each of these loops will contain 0, 1 or 2 qubits per code as required. A detailed description of this implementation and the shuttling schedule for each code will be discussed below. 

A useful feature of the linear-time CCZ procedure is that we only need to measure the $Z$ stabilisers of the slices and not the $X$ stabilisers (only the $Z$ stabiliser measurements are required to initialise the code and errors will not accumulate on the physical qubits because they are being constantly measured out and reinitialised). In fact, we could not measure the $X$ stabilisers even if we wished to because they do not commute with the CCZ gate, so they can be measured only in the first and last layers (at the start and end of the procedure) and not at any point in between. This is convenient because the $X$ stabilisers have fairly high weights and this results in a large amount of non-locality when mapped to 2D. The $Z$ stabilisers, on the other hand, are at most weight-4 and can be mapped to 2D much more easily. Accordingly, our pipelined implementation will allow for measurement of $X$ and $Z$ stabilisers in the layers but only for measurement of $Z$ stabilisers in the slices. 

\subsubsection{Layers}

\begin{figure}
    \centering
    \includegraphics[width=.48\textwidth]{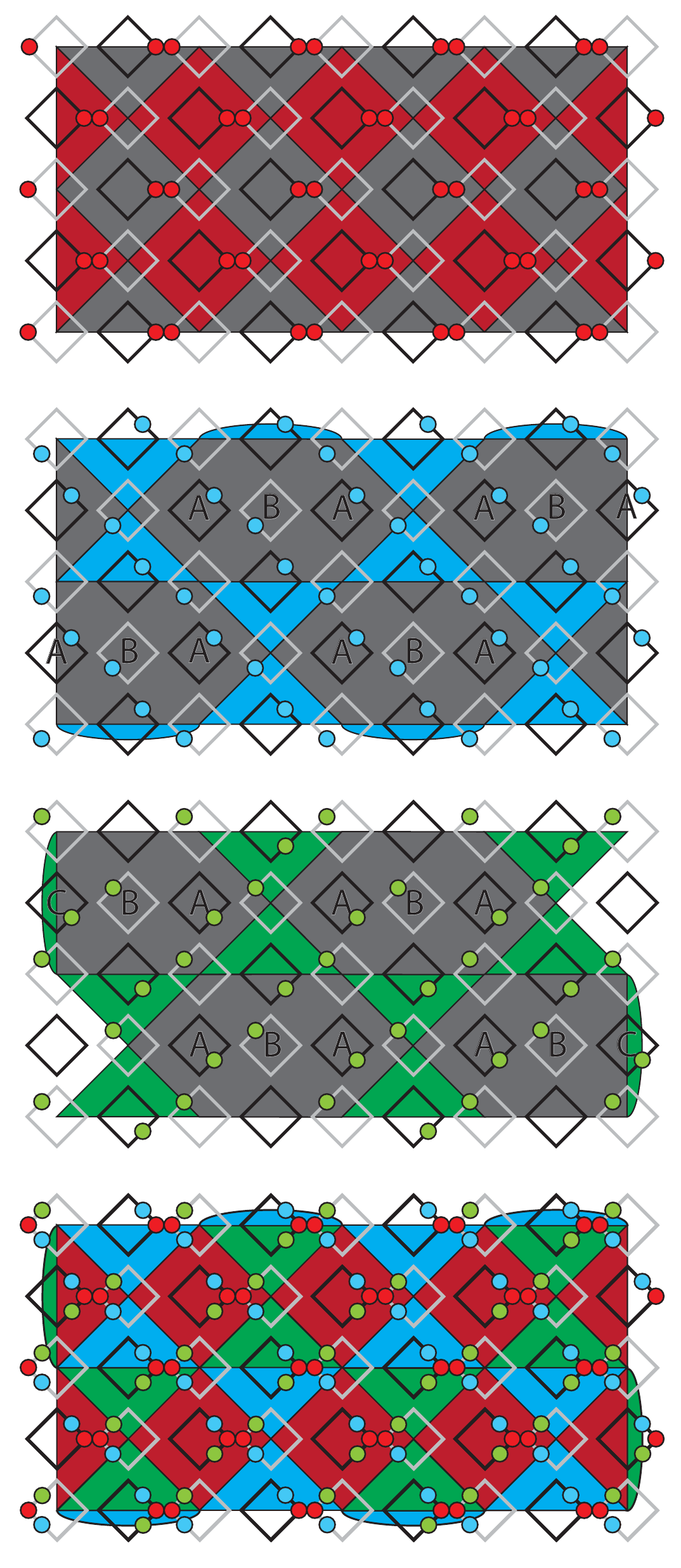}
    \caption{Pipelined implementation of the three 2D layers from \cref{fig:slices} and \cref{fig:grids}. Coloured (grey) faces are $Z$ ($X$) stabilisers. Grey loops contain data qubits and black loops contain ancilla qubits. Qubits move clockwise in all loops. The final image shows the overlap of all three layers. Only $Z$ stabilisers are shown in this case as this overlap only occurs at the midpoint of the procedure when we do not need to measure $X$ stabilisers.}
    \label{fig:loops_layers}
\end{figure}

\cref{fig:loops_layers} shows the pipelined implementation for each of the three layers. The first layer is simply the familiar 2D surface code on a square lattice and the implementation we obtain for this code is identical to the one described in~\cite{caiLoopedPipelinesEnabling2023}. The second and third images in \cref{fig:loops_layers} show implementations of the two kagome lattice layers. The most natural implementation these layers would measure each hexagonal $X$ stabiliser using a single large ancilla loop, the qubits of which could interact with all six data qubits in the stabiliser's support. However, we wish to use the same set of loops for all three codes so instead we measure these stabilisers using a two-step process. For the blue code (and in the bulk of the green code) we have two types of ancilla loop for each $X$ stabiliser, which we label \textbf{A} and \textbf{B} (notice that the \textbf{B} loops are ancilla qubit loops in these codes but data qubit loops in the red code). To measure the stabiliser the qubits in the two \textbf{A} loops are first entangled with all three adjacent data qubits, then they are both entangled with the qubit of the \textbf{B} loop which can then be measured out to complete the stabiliser measurement. At the $Z$ boundaries of the green code we have an additional loop type (\textbf{C}) whose ancilla qubit is responsible for measuring both a $Z$ boundary stabiliser and two of the qubits of an $X$ stabiliser. This is not a problem provided we first measure all $Z$ stabilisers and then all $X$ stabilisers rather than all in parallel (this is actually required even for the red code~\cite{caiLoopedPipelinesEnabling2023}). 

\subsubsection{Slices}\label{sec:slices}

\begin{figure}
    \centering
    \includegraphics[width=.48\textwidth]{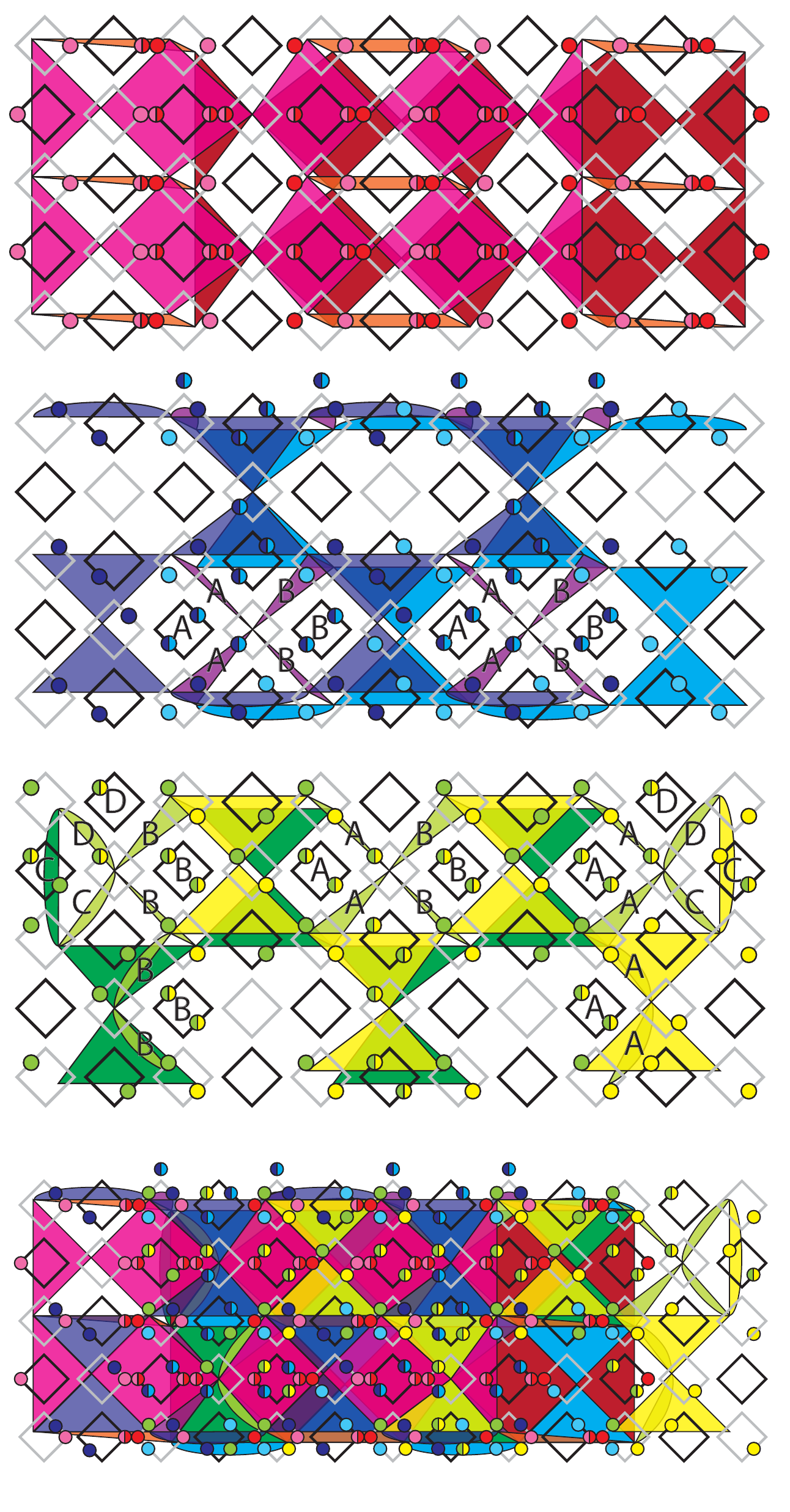}
    \caption{Pipelined implementation of the three slices from \cref{fig:slices,fig:grids}. The three colours of faces in each slice distinguish $Z$ checks that exist in the bottom layer, top layer, and only in the slice. Single-coloured (two-coloured) circles are data qubits belonging to only one layer (neither/both layers) or ancilla qubits for intra-layer (inter-layer) stabilisers.}
    \label{fig:loops_slices}
\end{figure}

As mentioned above, we do not need to measure the $X$ stabilisers of the slices so our proposed implementation (shown in \cref{fig:loops_slices}) only allows for $Z$ stabiliser measurements. Unlike in the layers, some of the loops now contain two qubits (one from each layer in the slice). In each of the three codes there are two different kinds of $Z$ stabiliser which we refer to as intra-layer and inter-layer stabilisers, with the former being supported entirely on qubits from the same layer and the latter supported on qubits from both layers, or on qubits that exist in the slice but in neither layer. All data-ancilla interactions necessary for measuring the intra-layer stabilisers can be performed in a single loop cycle. However, the ancilla qubits for the inter-layer stabilisers are in phase with the data qubits from only one of the layers, but in antiphase with the data qubits of the other layer. We therefore propose the following three-step procedure for measuring all the stabilisers of these slices

\begin{enumerate}
    \item Perform one full loop cycle. All interactions for intra-layer stabilisers and all possible interactions for inter-layer stabilisers are performed. 
    \item Advance all inter-layer ancilla by half a cycle.
    \item Perform one more full loop cycle. The remaining operations for the inter-layer stabilisers can be performed in this cycle.
\end{enumerate}

The operations performed in step $1$ are similar to that of a code cycle in the 2D codes. Step 2 consists of pure shuttling, which is usually not the rate-limiting step in many hardware~\cite{kaushalShuttlingbasedTrappedionQuantum2020,langrockBlueprintScalableSpin2023}. The number of operations performed in step $3$ is fewer than the usual code cycles since we are only interacting with the data qubits in one of the layers for the inter-layer stabilisers. Hence, we will approximate the total time needed in steps 2 and 3 as one code cycle, which means the full process of the three steps above takes around two code cycles.

Because the partitioning of loops into ``data'' and ``ancilla'' is consistent across all three slices (unlike with the layers) the half-cycle advancement of the ancilla qubits does not interfere with the scheduling of any other qubits. Even in the case where intra- and inter-layer ancillas share a loop there are no issues because the intra-layer ancillas have already completed all necessary interactions. 

For the first image in \cref{fig:loops_slices} the correspondence between inter-layer stabilisers and loops is straightforward as each loop containing one (two) inter-layer ancilla overlaps with exactly one (two) of the inter-layer stabilisers. This is less straightforward in the second and third images and additional modifications are required to accommodate the boundary stabilisers in these codes. In \cref{fig:loops_slices} the inter-layer ancillas in the loops labelled \textbf{A} measure the two inter-layer stabilisers (also labelled \textbf{A}) to their right while those labelled \textbf{B} measure the two inter-layer stabilisers to their left. Additionally, the green code has loops labelled \textbf{C} and \textbf{D} which each contain one inter-layer ancilla that measures the adjacent stabiliser with that same label. The \textbf{C} loops also contain an intra-layer ancilla that measures the intra-layer stabiliser which overlaps with the loop. Finally, one boundary of the blue code supports boundary stabilisers which are supported on two qubits of the same loop. In isolation we could measure these stabilisers by interacting these two qubits with each other or by adding an ancilla qubit to this loop, but when multiple slices overlap these two qubits will be separated by qubits from the other codes. Instead, we introduce static ancilla qubits next to this boundary which can interact with the data qubits as they move past.

In the last image we show the overlap of all three slices. Notice that in the loops which contain six qubits these qubits are arranged in two groups of three so that the same-layer qubits from all three codes are together. These are the triples of qubits between which we need to apply CCZ, so this arrangement makes the application of this gate more straightforward. 

\subsection{In-Place Gate}\label{sec:in_place_gate}

\begin{figure}
    \centering
    \includegraphics[width=.48\textwidth]{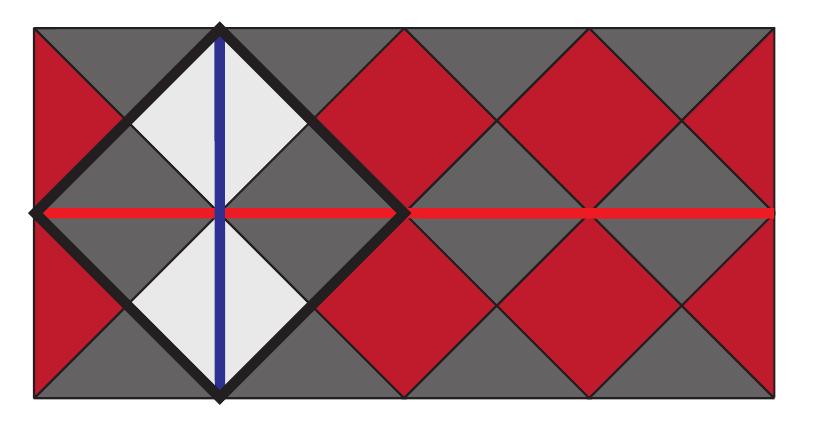}
    \caption{Conversion of a distance-3 rotated surface code (thick black lines) to the red code shown previously. New $Z$ stabilisers are shown in red. $X$ (blue) and $Z$ (red) logical operator representations are also shown.}
    \label{fig:lattice_conversion1}
\end{figure}

\begin{figure}
    \centering
    \includegraphics[width=.48\textwidth]{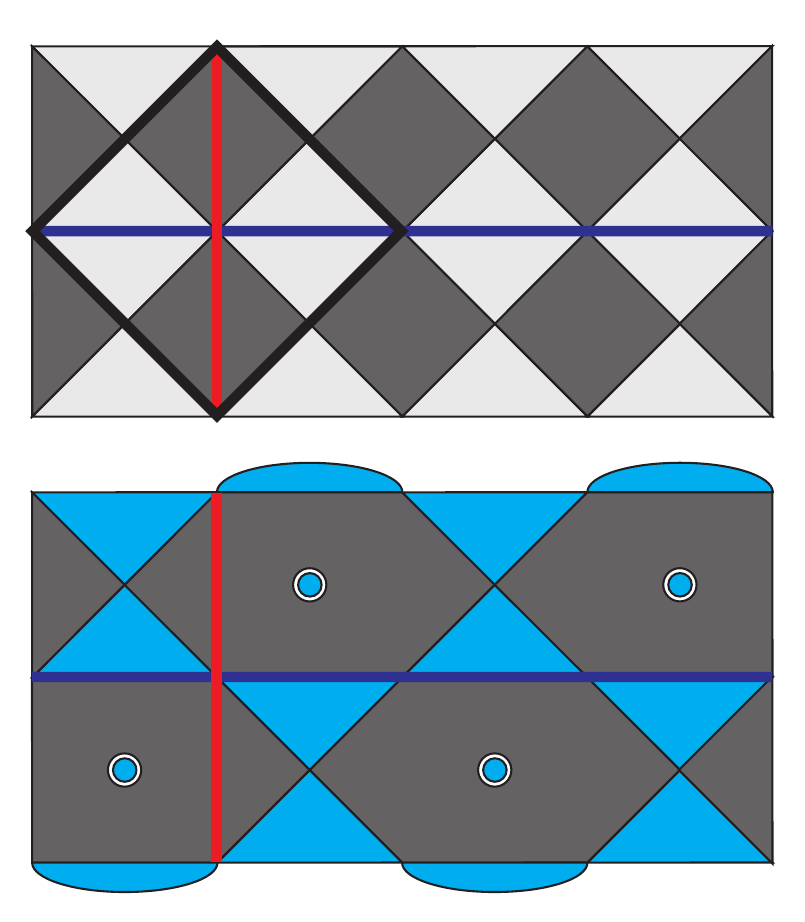}
    \caption{Conversion of a distance-3 rotated surface code to the blue code shown previously. The code is first expanded to an unrotated code and then some qubits are measured out in the $Z$ basis (blue circles). $X$ checks in the new code are products of checks from the old code, while $Z$ checks are checks from the old code restricted to their supports on unmeasured qubits. The displayed logical operator implementations are valid for both initial and final codes.}
    \label{fig:lattice_conversion2}
\end{figure}

\begin{figure}
    \centering
    \includegraphics[width=.48\textwidth]{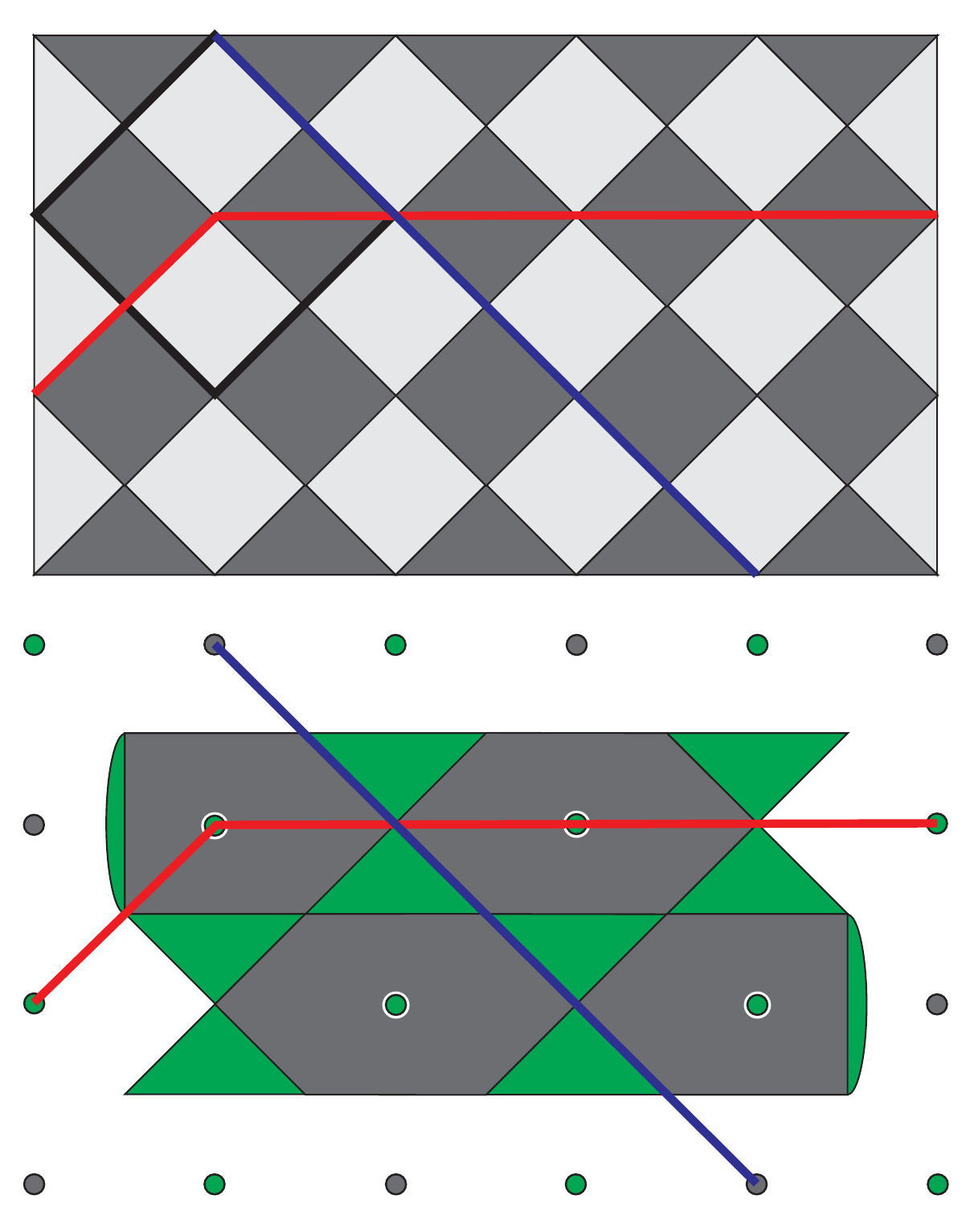}
    \caption{Conversion of a distance-3 rotated surface code to the green code shown previously. The code is first expanded to a large unrotated code and then some qubits are measured out in either the $X$ or $Z$ basis (green for $Z$ and grey for $X$). $X$ and $Z$ checks in the new code are (products of) checks in the old code restricted to their supports on unmeasured qubits. The displayed logical operator implementations commute with all new stabilisers and single-qubit measurements.}
    \label{fig:lattice_conversion3}
\end{figure}

Besides using the linear-time CCZ gate as a way to prepare magic states without using distillation, it can also perform logical CCZ directly between any triple of surface code lattice patches in the computer by moving these patches into the correct positions and using code deformation techniques to transform the lattices as necessary. These transformations are shown in \cref{fig:lattice_conversion1}, \cref{fig:lattice_conversion2}, and \cref{fig:lattice_conversion3}. The transformation is simplest for the red code (\cref{fig:lattice_conversion1}) where we just need to prepare new qubits in $\ket{+}$ and measure additional $Z$ stabilisers to transform a rotated surface code into the unrotated code required for the gate. For the blue code (\cref{fig:lattice_conversion2}) we show the transformation in two steps, first the rotated code is transformed into an unrotated code and then single-qubit $Z$ measurements are used to transform from a square lattice to a kagome lattice. For the green code the process is similar, but the rotated code must first be transformed into a large unrotated code, then single-qubit $X$ and $Z$ measurements are used to obtain the target code. 

Implementing logical CCZ using this in-place gate instead of magic state injection removes the cost associated with routing magic states from factories to the places they will be consumed, and also avoids computational bottlenecks that may occur if the rate of magic state consumption is higher than the rate of production.

\begin{figure*}
    \centering
    \subfloat[]{
        \includegraphics[width=.23\textwidth]{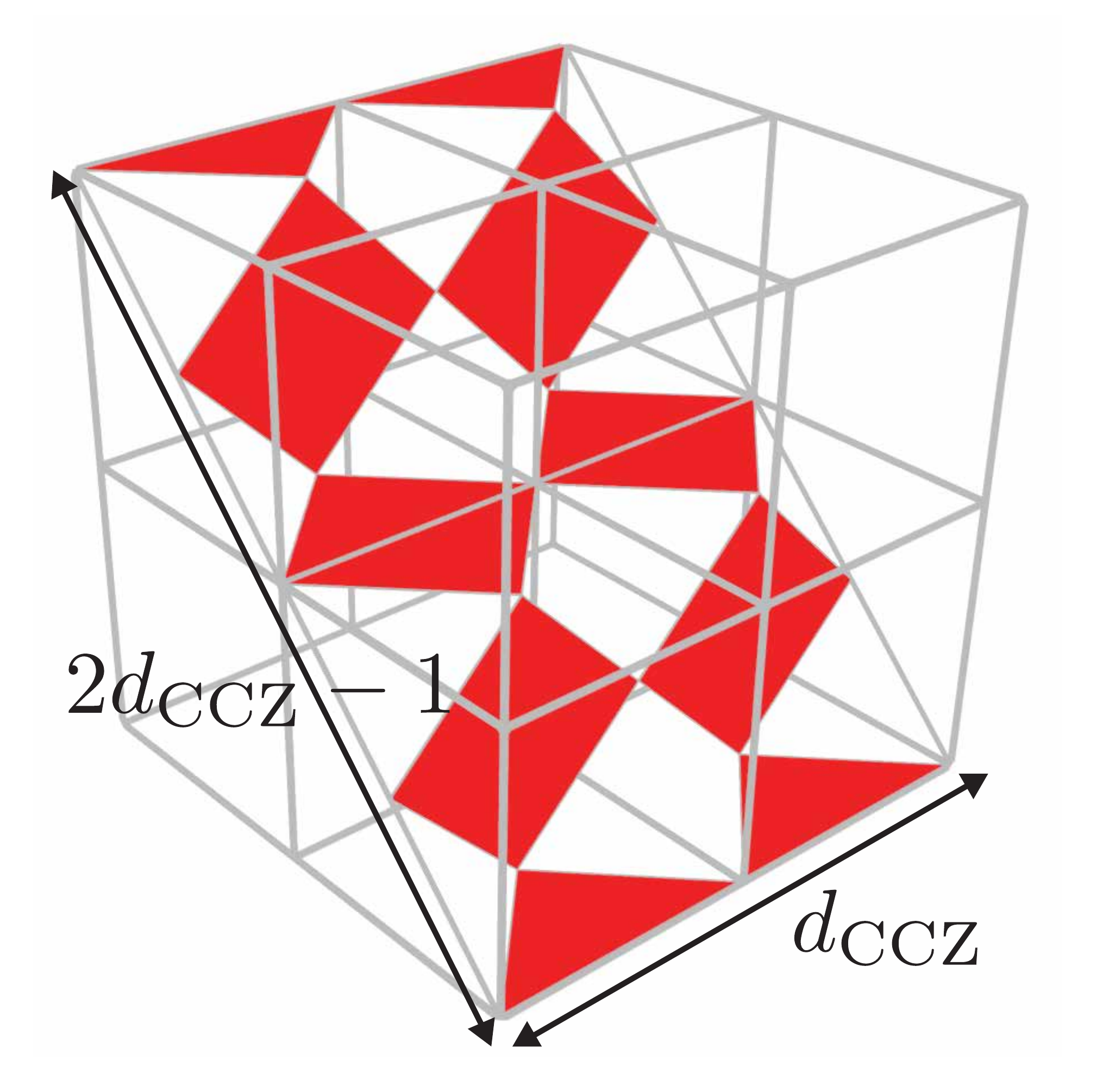}
        }
    \subfloat[]{
        \includegraphics[width=.73\textwidth]{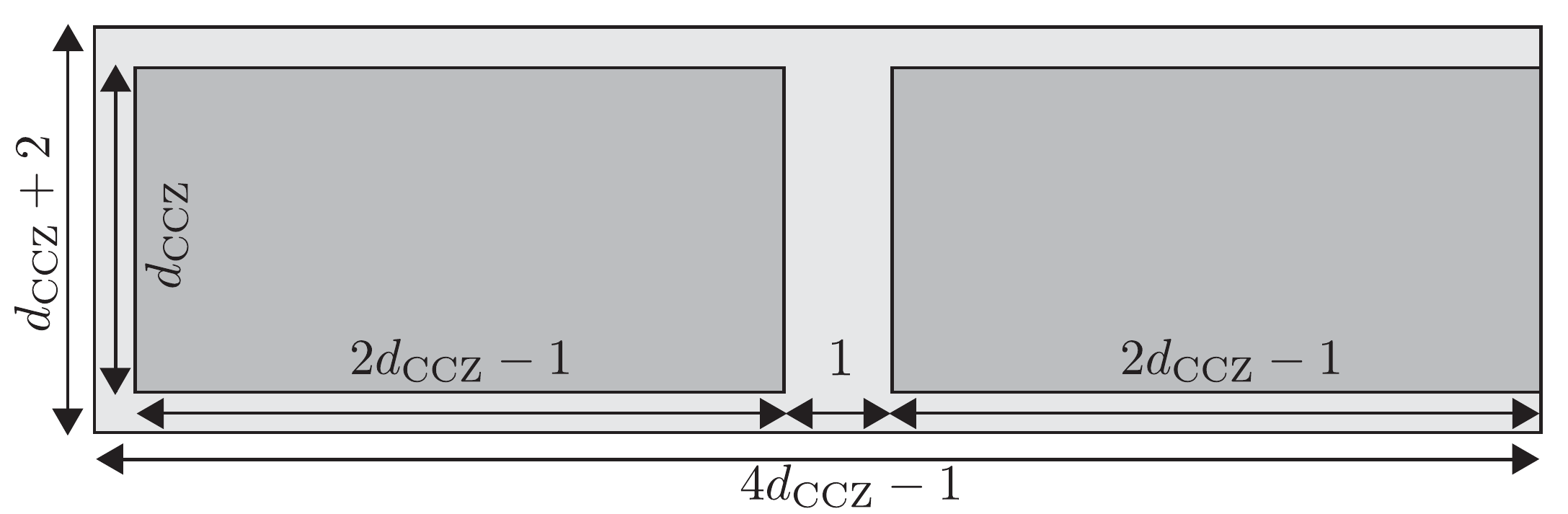}
        }
    \caption{\textbf{(a)} A distance $d_\textsc{ccz}$ layer of one code (here the red code) fits inside a $d_\textsc{ccz} \times d_\textsc{ccz} \times d_\textsc{ccz}$ cube using $2d_\textsc{ccz}-1$ data qubits on the diagonal edge (here $d_\textsc{ccz}=3$). \textbf{(b)} Code layout (measured in number of data qubit loops) needed for linear CCZ gates. The dark grey regions represent the space required for the three codes, while the extra space in the light grey region is used for the transformation of the green code from rotated to unrotated and for some of the boundary ancillas in the blue code. Here we assume that the green code is on the left while the red and blue codes are on the right.} 
    \label{fig:measuring}
\end{figure*}

We will write the distance of the codes used in the in-place gate as $d_\textsc{ccz}$, and note that in general this can differ from the distance used when performing Clifford operations with the rotated codes. We can count the number of data qubit loops required to implement this gate by first noticing that a single layer with distance $d_\textsc{ccz}$ fits inside a $d_\textsc{ccz} \times d_\textsc{ccz} \times d_\textsc{ccz}$ cube, as shown in \cref{fig:measuring}(a). There are then $d_\textsc{ccz}$ data qubits along the horizontal edge of the layer and $2d_\textsc{ccz}-1$ qubits on the diagonal edge ($d_\textsc{ccz}$ qubits on grid vertices plus $d_\textsc{ccz}-1$ qubits on grid faces). As mentioned in \cref{subsection:JIT_background}, the target code patches of the CCZ gates start in two disjoint stacks, slide over each other during the application of the gate and end in disjoint stacks again. Hence, to perform the CCZ gate we need space to place two stacks end-to-end with a small space (at least one data qubit loop) between them, as shown in \cref{fig:measuring}(b). We also need to have a small amount of extra space (two loops in one direction and one in the other) around the edges of these patches for the transformation of the green code from rotated to unrotated, and for the boundary ancillas for the blue code. The final dimensions are then $(4d_\textsc{ccz}-1) \times (d_\textsc{ccz}+2)$ data qubit loops.

The full CCZ requires the codes on the left and right to switch places, so each code needs to travel $(2d_\textsc{ccz}-1) + 1 = 2d_\textsc{ccz}$ data qubit loops to the left or right. Each step of the gate consists of three stages: expanding a layer to a slice, applying transversal physical CCZ and collapsing to the next layer, and the code will travel to left or right by one data qubit loop after one step. Thus, in total $2d_\textsc{ccz}$ steps are required for the full CCZ gate. The physical CCZ gate can be compiled using $6$ CNOTs as outlined in \cite{nielsenQuantumComputationQuantum2010}. Alternatively, there can be native implementations of CCZ (or Toffoli) in different platforms~\cite{gullansProtocolResonantlyDriven2019,liQuantumFredkinToffoli2022,goelNativeMultiqubitToffoli2021,fedorovImplementationToffoliGate2012,kimHighfidelityThreequbitIToffoli2022}. For simplicity, we will assume the gate error rate for CCZ is the same as the two-qubit error rate, since we are trying to see if linear CCZ can compete with magic state distillation using the most favourable assumption. As discussed in \cref{sec:slices}, the expand stage requires $2$ code cycles. The collapse stage is simply measuring out the relevant data qubits, combined with the gate time needed for the physical CCZ, the total time needed for both of these stages is roughly $1$ code cycle. In this way, the total number of code cycles needed per step is $2 + 1 = 3$, and the total number of code cycles needed for performing linear CCZ is $T_{\mathrm{lin}} = 3 \times 2d_\textsc{ccz} = 6d_\textsc{ccz}$.

